\documentclass[12pt]{article}

\usepackage{graphicx}
\graphicspath{{figures/}} 
\usepackage{caption}
\usepackage{subcaption}
\usepackage{physics}
\usepackage{array,longtable} 

\usepackage{amssymb}
\usepackage{amsmath}
\usepackage{dsfont}				
\usepackage{slashed}			
\usepackage[compat=1.0.0]{tikz-feynman}			

\usepackage{jheppub}
\usepackage{orcidlink} 

\newcommand{\mbbone}{\text{\usefont{U}{bbold}{m}{n}1}}
\MakeRobust{\mbbone}

\def\a{\alpha}
\def\b{\beta}

\def\const{\mathrm{const}}

\def\d{\delta}
\def\D{\Delta}

\def\eps{\varepsilon}
\def\f{\frac}

\def\l{\left}
\def\la{\langle}
\def\ra{\rangle}

\def\mc{\mathcal}

\def\nn{\nonumber}

\def\r{\right}

\def\t{\theta}

\def\be{\begin{equation}}
\def\ee{\end{equation}}
\def\bea{\begin{eqnarray}}
\def\eea{\end{eqnarray}}
\def\ba{\begin{array}}
\def\ea{\end{array}}
\def\bc{\begin{center}}
\def\ec{\end{center}}
\def\bl{\begin{flushleft}}
\def\el{\end{flushleft}}
\def\br{\begin{flushright}}
\def\er{\end{flushright}}
\def\bi{\begin{itemize}}
\def\ei{\end{itemize}}
\def\bt{\begin{tabular}}
\def\et{\end{tabular}}

\title{Spinning the Large-Charge Bootstrap: Parity-Even Operators}

\author{Kasra Kiaee~\orcidlink{0009-0001-9383-9491},}
\author{Alexander Monin~\orcidlink{0000-0001-9219-6474}}

\affiliation{University of South Carolina \\
712 Main St, 404 \\
Columbia SC, 29208

}
\emailAdd{kkiaee@email.sc.edu}
\emailAdd{amonin@mailbox.sc.edu}

\abstract{
We study the large-charge bootstrap in three-dimensional CFTs with a global \(U(1)\) symmetry using scalar, current, and stress-tensor probes, restricting to parity-even exchanged operators. Under minimal assumptions, the bootstrap requires at least one Regge trajectory with the dispersion relation of the standard Goldstone mode of the conformal superfluid. With additional assumptions, this trajectory is unique, while all non-Goldstone trajectories contribute only to the scalar-scalar channel at this order.
}

\notoc
\begin{document}

\maketitle

\newpage

\tableofcontents

\section{Introduction}

Conformal field theories (CFTs) are highly constrained by symmetry, but they remain difficult to solve in general. Conformal invariance fixes the kinematics to a remarkable degree. Correlation functions of primary operators are restricted to a finite number of tensor structures which can be constructed systematically using the embedding-space methods of~\cite{Costa:2011mg,Costa:2011dw}.

The real difficulty is dynamical. A CFT is specified by its spectrum of scaling dimensions and by its operator product expansion (OPE) coefficients. These data are only weakly constrained by symmetry alone, and determining them is hard in general. There are important examples where general nonperturbative results can be obtained, most notably through the numerical conformal bootstrap, which has led to very precise determinations of CFT data in the three-dimensional Ising model~\cite{Simmons-Duffin:2016wlq}. In a generic CFT there is no obvious small parameter, no obvious large parameter, and no simple perturbative expansion. Without extra assumptions or additional organizing principles, extracting dynamical information appears to be an impasse.

A major success was the analytic conformal bootstrap at large spin~\cite{Fitzpatrick:2012yx,Komargodski:2012ek}. In this regime, crossing symmetry can be used systematically, leading in particular to the prediction of infinite families of operators with arbitrarily large spin and asymptotically identical twists. Thus the large-spin limit provides an example where a universal sector of a generic CFT can be accessed without knowing the full theory.

A different universal regime arises in sectors of large global charge. For CFTs with a global symmetry, it was proposed in~\cite{Hellerman:2015nra,Monin:2016jmo} that in many cases the lowest-dimension operator of large charge \(Q\) admits an effective field theory description. By the operator-state correspondence, this operator defines the lowest-energy state of charge \(Q\) on the cylinder. In the standard realization, this state is a superfluid, and the low-energy dynamics is governed by the associated Goldstone mode. This picture is quite general, but it is not automatic. Some theories, such as the fermionic examples discussed in~\cite{Dondi:2022zna}, are described instead by a Fermi liquid rather than by a superfluid.

The first attempt to understand how generic the superfluid EFT prediction is from CFT consistency alone was made in~\cite{Jafferis:2017zna}. There the authors considered four-point functions with two large-charge scalar operators and two light scalar probes. Under mild assumptions, they derived a general solution to the crossing equations at leading and first subleading order in the large-\(Q\) expansion. In particular, they obtained the allowed scaling dimensions and OPE coefficients of the states that contribute at this order.

This construction was an important step, but it was not restrictive enough to determine under what conditions the superfluid EFT exhausts the possible large-charge dynamics. The reason is that scalar probes do not make use of operators that are necessarily present in any CFT with a global \(U(1)\) symmetry. A natural improvement was made in~\cite{Kiaee:2025jly}, where the scalar bootstrap equations were enlarged to a coupled system including the conserved \(U(1)\) current as a probe.

The inclusion of the current led to new constraints on the allowed solutions within the parity-even sector considered there. In particular, the bootstrap equations imply the existence of a Regge trajectory whose analytically continued energy vanishes at spin \(\ell=0\). Nevertheless, even under the assumptions of that analysis, the result did not fully determine whether the allowed solutions are necessarily described by the superfluid EFT, possibly supplemented by additional light fields, or whether genuinely different solutions to the large-charge bootstrap equations exist.

In this paper we consider three-dimensional CFTs with a global \(U(1)\) symmetry.\footnote{Some formulas that can be derived in general dimension \(d\), and do not require additional assumptions, are presented for arbitrary \(d\).} We study the crossing equations for four-point functions involving two large-charge scalar operators and two light probes. The probes can be scalar operators, the conserved \(U(1)\) current \(J^a\), or the stress tensor \(T^{ab}\). At the order in the large-charge expansion considered here, we restrict the \(s\)- and \(u\)-channel OPEs to parity-even exchanged operators. The goal is to understand how much of the large-charge dynamics in this case is fixed by crossing, unitarity, and the existence of these universal operators.

We adopt the following minimal assumptions about the CFT:

\begin{enumerate}
\item
For every \(Q\), the lowest-dimension operator in the charge-\(Q\) sector is assumed to be a unique scalar primary, which we denote by \(\Phi_Q\), with scaling dimension \(\Delta_Q\). Equivalently, the charge-\(Q\) ground state on the cylinder is non-degenerate and is a scalar.

\item
The macroscopic limit exists. Namely, we put the theory on the cylinder \(\mathbb R\times S^{d-1}\) and take
\begin{equation}
Q\to\infty,
\qquad
R\to\infty,
\end{equation}
while keeping both the charge density and the energy density fixed. This limit is also discussed in Section~\ref{subsec:macroscopic-limit}. In particular, it implies the scaling
\begin{equation}
\Delta_Q=\alpha |Q|^{\frac{d}{d-1}},
\qquad
\alpha=\const.
\end{equation}

\item
At first subleading order in the large-\(Q\) expansion, the \(s\)-channel contains two types of contributions: the first descendant of the leading state \(\Phi_Q\), and new primary states that enter at the same order in \(1/Q\).

\item
Only a finite number of Regge trajectories contribute to the \(s\)-channel expansion at this order.

\item
At the order considered here, only parity-even operators contribute to the \(s\)- and \(u\)-channel OPEs.
\end{enumerate}
Under these minimal assumptions, we show that the spectrum necessarily contains at least one Regge trajectory corresponding to the standard Goldstone mode of the conformal superfluid. In three dimensions, this trajectory has sound speed
\begin{equation}
c_s^2=\frac{1}{2}.
\end{equation}

A reader familiar with large-charge CFTs may notice that this conclusion appears to be in tension with the spectrum of the conformal solid discussed in~\cite{Esposito:2017qpj}. The conformal solid, however, lies outside the class of solutions covered by our minimal assumptions. Indeed, in the conformal solid, parity-odd operators related to the transverse phonons contribute to the \(s\)- and \(u\)-channel OPEs at the order considered here, whereas our analysis is restricted to parity-even exchanged operators \cite{Cuomo:2021qws}. The extension of the analysis to parity-odd exchanged operators will be presented elsewhere.

In parts of the analysis we impose the following additional assumptions. They are not required for deriving the crossing constraints themselves and should in no sense be regarded as necessary. They are introduced only to sharpen the interpretation of particular results:

\begin{enumerate}

\setcounter{enumi}{5}

\item
\label{item:SpinZeroContinuation}
We assume that no non-Goldstone Regge trajectory reaches zero energy when analytically continued to spin \(\ell=0\). Under this assumption, we show that the only states that can couple to the current \(J\) belong to the Goldstone trajectory. Therefore, all non-Goldstone trajectories, if present, can contribute only to the scalar-scalar channel. This is precisely the structure expected from a large-charge EFT in which the universal Goldstone mode is accompanied by additional light fields.

\item
\label{item:NonDegeneracy}
We assume non-degeneracy of the charge-\(Q\) spectrum at fixed spin at the order in the large-\(Q\) expansion considered here. Assuming non-degeneracy only among the new primary states entering at this order implies that the conformal-superfluid Goldstone trajectory is unique and allows us to isolate the contribution of at most one physical spin-one primary with excitation energy \(1\), rather than constraining only a linear combination of the fusion coefficients of all such primaries. A stronger version of the assumption excludes degeneracy specifically between the first descendant of the ground-state primary and the new spin-one primary entering at this order. It therefore excludes the additional spin-one primary with excitation energy \(1\).

\end{enumerate}
These additional assumptions are used only to obtain more restrictive conclusions and are not part of the minimal setup.

The paper is organized as follows. In Section~\ref{sec:setup}, we introduce our conventions for four-point functions with scalar, current, and stress-tensor probes, and construct the tensor structures used throughout the analysis. In Section~\ref{sec:large-charge-expansion}, we derive the leading and next-to-leading contributions to the six correlator sectors
\begin{equation}
SS,\qquad SJ,\qquad JJ,\qquad ST,\qquad JT, \qquad TT,
\end{equation}
including the contributions of the leading charge-\(Q\) state, its first descendant, and the new primary states entering at this order. In Section~\ref{sec:bootstrap-equations}, we formulate the corresponding crossing equations. The existence of the macroscopic limit determines the behavior of the correlators near the \(t\)-channel singularity at \(z=1\), while smoothness away from the singularity allows us to rewrite crossing as an infinite set of algebraic constraints on the spectral data.

In Section~\ref{sec:solutions}, we analyze these constraints. The \(TT\) sector implies that any trajectory coupling to the stress tensor must have
\begin{equation}
\omega_\ell^2
=
J_\ell^2
\equiv
\frac{\ell(\ell+1)}{2},
\end{equation}
and consistency requires that at least one such trajectory be present. We then study the remaining scalar and current sectors and determine how the conclusions are strengthened under the additional assumptions stated above. We conclude in Section~\ref{sec:conclusions}. Technical details, including the relation between the tensor bases, conservation constraints, the full crossing equations, and supplementary proofs, are collected in the appendices.

\section{Setup}
\label{sec:setup}

\subsection{Conventions}

Although some intermediate formulas will be written for arbitrary \(d\), in this paper we will only be interested in \(d=3\). The map between \(d\)-dimensional Euclidean space \(\mathbb R^d\) and the cylinder
\(\mathbb R \times \mathbb S^{d-1}\), with coordinates \((\tau,\t^i)\), is realized by\footnote{Everywhere we set the radius of the sphere to be \(R=1\).}
\begin{equation}
\label{eq:CylinderMap}
x^a = e^{\tau} n^a(\t), \qquad n^a n^a = 1 .
\end{equation}
Given a local operator \({\mc O}^{a_1\dots}_{b_1\dots}(x)\) in Euclidean space with scaling dimension \(\D_{\mc O}\), the corresponding operator on the cylinder is defined as
\begin{equation}
\label{eq:OperatorCylinder}
\hat {\mc O}^{a_1\dots}_{b_1\dots}(\tau,\vec n)
=
|x|^{\D_{\mc O}+\#a-\#b}
{\mc O}^{a_1\dots}_{b_1\dots}(x),
\end{equation}
where \(\#a\) and \(\#b\) denote the number of upper and lower indices. We also introduce the rescaled operator
\begin{equation}
\label{eq:OperatorRescaled}
\tilde {\mc O}^{a_1\dots}_{b_1\dots}(\tau,\vec n)
=
|x|^{\D_{\mc O}}
{\mc O}^{a_1\dots}_{b_1\dots}(x).
\end{equation}
For scalar operators the two definitions coincide.

States on the cylinder are defined via the operator-state correspondence
\begin{equation}
\label{eq:OperatorState}
|{\mc O}\ra
=
\lim_{x\to 0}
{\mc O}(x)|0\ra .
\end{equation}
For a scalar primary, Hermitian conjugation is given by
\begin{equation}
\label{eq:ScalarHermitian}
\la \phi |
=
\lim_{x\to \infty}
x^{2\D}
\la 0|\phi(x),
\end{equation}
and similarly for a spin \(\ell\) operator, namely a rank-\(\ell\) traceless symmetric tensor,
\begin{equation}
\label{eq:SpinHermitian}
\la T^{a_1\dots a_\ell} |
=
\lim_{x\to \infty}
x^{2\D}
I^{a_1 b_1}\dots I^{a_\ell b_\ell}
\la 0|T^{b_1\dots b_\ell}(x),
\end{equation}
with
\begin{equation}
\label{eq:InversionTensor}
I^{ab}
=
\d^{ab}-2n^a n^b .
\end{equation}

We define the conformal cross-ratios
\begin{equation}
\label{eq:CrossRatios}
u
=
\f{x_{12}^2 x_{34}^2}{x_{13}^2 x_{24}^2},
\qquad
v
=
\f{x_{14}^2 x_{23}^2}{x_{13}^2 x_{24}^2}.
\end{equation}
We will also trade \((u,v)\) for the cylinder coordinates \((\tau,\t)\), where
\begin{align}
\label{eq:CylinderCoordinates}
\tau
&=
\f{1}{2}\log u,
\\
\t
&=
\arccos
\f{1+u-v}{2\sqrt{u}} .
\end{align}

\subsection{$H$-basis}

We are interested in four-point functions on the plane involving two lowest large-charge operators and two light probes. The scalar probe is taken to be charged. In the scalar-scalar sector we consider
\begin{equation}
\label{eq:PlaneFourPointSS}
\left\langle
\Phi_{-Q}(x_4)\,
\phi_{-q}(x_3)\,
\phi_q(x_2)\,
\Phi_Q(x_1)
\right\rangle .
\end{equation}
For correlators involving one charged scalar probe and either the current or the stress tensor, we consider\footnote{
The asymmetric-looking assignment of charges in \eqref{eq:PlaneFourPointSA} is chosen for later convenience.}
\begin{equation}
\label{eq:PlaneFourPointSA}
\left\langle
\Phi_{-\left(Q-\frac q2\right)}(x_4)\,
\phi_{-q}(x_3)\,
{\cal A}(x_2)\,
\Phi_{Q+\frac q2}(x_1)
\right\rangle ,
\qquad
{\cal A}=J^a,\;T^{ab}.
\end{equation}
For correlators involving only neutral spinning probes, we consider
\begin{equation}
\label{eq:PlaneFourPointAB}
\left\langle
\Phi_{-Q}(x_4)\,
{\cal A}_3(x_3)\,
{\cal A}_2(x_2)\,
\Phi_Q(x_1)
\right\rangle ,
\qquad
{\cal A}_2,{\cal A}_3\in\{J^a,T^{ab}\}.
\end{equation}
The probe operators are therefore chosen from
\begin{equation}
\label{eq:ProbeOperators}
{\cal O}\in
\left\{
\phi_q,\,
\phi_{-q},\,
J^a,\,
T^{ab}
\right\},
\end{equation}
where \(\phi_{\pm q}\) are scalar probes of charge \(\pm q\), \(J^a\) is the conserved \(U(1)\) current, and \(T^{ab}\) is the energy-momentum tensor.

Taking the limit
\begin{equation}
x_4\to \infty,
\qquad
x_1\to 0,
\end{equation}
and using the operator-state correspondence, the four-point functions above can be recast as two-point functions on the cylinder. For the scalar-scalar sector we define
\begin{equation}
\label{eq:CylinderTwoPointSS}
|z|^{\Delta_Q}\,
G^{(Q)}_{SS}
=
\la Q|
\tilde \phi_{-q}(x_3)\,
\tilde \phi_q(x_2)
|Q\ra .
\end{equation}
For the mixed scalar-current and scalar-tensor sectors, we define
\begin{equation}
\label{eq:CylinderTwoPointSA}
|z|^{\frac12\left(\Delta_{Q-\frac q2}+\Delta_{Q+\frac q2}\right)}
G^{(Q)}_{S{\cal A}}
=
\left\langle Q-\frac q2\right|
\tilde \phi_{-q}(x_3)\,
\tilde{\cal A}(x_2)
\left|Q+\frac q2\right\rangle .
\end{equation}
Finally, for the sectors involving only neutral spinning probes, we define
\begin{equation}
\label{eq:CylinderTwoPointAA}
|z|^{\Delta_Q}\,
G^{(Q)}_{{\cal A}_3{\cal A}_2}
=
\la Q|
\tilde{\cal A}_3(x_3)\,
\tilde{\cal A}_2(x_2)
|Q\ra ,
\qquad
{\cal A}_2,{\cal A}_3\in\{J^a,T^{ab}\}.
\end{equation}
Here the tilde denotes the rescaled operator defined in
\eqref{eq:OperatorRescaled}.

For scalar probes, the correlator \(G_{SS}\) is a function only of the conformal cross-ratios \(u\) and \(v\), or equivalently of the cylinder variables \(\tau\) and \(\t\). For correlators involving vector or tensor indices, the same statement is true only after decomposing the correlator into tensor structures. Namely, the full correlator is written as a sum of independent tensor structures, with coefficient functions depending on \(u\) and \(v\). There are several equivalent choices of tensor basis. We present two of them in Appendices~\ref{app:F-basis} and~\ref{app:F-to-H}. The \(H\)-basis, described in Appendix~\ref{app:F-to-H}, makes the crossing equations and the subsequent algebraic reduction more transparent.

The \(H\)-basis functions are constructed from the correlators by taking suitable contractions with the unit vectors \(n_2^a\) and \(n_3^a\). We use
\begin{equation}
n_2^a n_2^a = n_3^a n_3^a = 1,
\qquad
n_2^a n_3^a = \cos\t .
\end{equation}
The word ``level'' denotes the number of contractions needed to obtain the corresponding structure. Thus level zero terms are read directly from the correlator, while higher-level terms are obtained after contracting one or more indices with \(n_2^a\) or \(n_3^a\). When we write \(\supset\), we only display the tensor structure whose coefficient defines the corresponding \(H\)-function.

The signs in front of the coefficient functions below are chosen for later convenience.

\paragraph{Scalar-scalar}

There is only one structure, and therefore
\begin{equation}
G^{(Q)}_{SS}=H^{(Q)}_{SS}.
\end{equation}

\paragraph{Scalar-current}

\begin{equation}
\begin{alignedat}{3}
\text{Level 0}: \quad&
G^{(Q)}_{SJ,a}
&\quad&\supset
&\quad& H^{(Q)}_{SJ,3}\,n_3^a,
\\
\text{Level 1}: \quad&
G^{(Q)}_{SJ,a}n_2^a
&\quad&=
&\quad& H^{(Q)}_{SJ,0}.
\end{alignedat}
\end{equation}

\paragraph{Current-current}

\begin{equation}
\begin{alignedat}{3}
\text{Level 0}: \quad&
G^{(Q)}_{JJ,ab}
&\quad&\supset
&\quad&
H^{(Q)}_{JJ,23}\,n_2^a n_3^b
+
H^{(Q)}_{JJ,\delta}\,\delta^{ab},
\\
\text{Level 1}: \quad&
G^{(Q)}_{JJ,ab}n_3^a
&\quad&\supset
&\quad&
- H^{(Q)}_{JJ,3}\,n_3^b,
\\
&
G^{(Q)}_{JJ,ab}n_2^b
&\quad&\supset
&\quad&
H^{(Q)}_{JJ,2}\,n_2^a,
\\
\text{Level 2}: \quad&
G^{(Q)}_{JJ,ab}n_3^a n_2^b
&\quad&=
&\quad&
- H^{(Q)}_{JJ,0}.
\end{alignedat}
\end{equation}

\paragraph{Scalar-tensor}

\begin{equation}
\begin{alignedat}{3}
\text{Level 0}: \quad&
G^{(Q)}_{ST,ab}
&\quad&\supset
&\quad&
- H^{(Q)}_{ST,33}\,n_3^a n_3^b,
\\
\text{Level 1}: \quad&
G^{(Q)}_{ST,ab}n_2^b
&\quad&\supset
&\quad&
H^{(Q)}_{ST,3}\,n_3^a,
\\
\text{Level 2}: \quad&
G^{(Q)}_{ST,ab}n_2^a n_2^b
&\quad&=
&\quad&
H^{(Q)}_{ST,0}.
\end{alignedat}
\end{equation}

\paragraph{Current-tensor}

\begin{equation}
\begin{alignedat}{3}
\text{Level 0}: \quad&
G^{(Q)}_{JT,abc}
&\quad&\supset
&\quad&
- H^{(Q)}_{JT,233}\,n_2^a n_3^b n_3^c
-
H^{(Q)}_{JT,\delta3}
\left(
\delta^{ab}n_3^c+\delta^{ac}n_3^b
\right),
\\
\text{Level 1}: \quad&
G^{(Q)}_{JT,abc}n_2^c
&\quad&\supset
&\quad&
H^{(Q)}_{JT,23}\,n_2^a n_3^b
+
H^{(Q)}_{JT,\delta}\,\delta^{ab},
\\
&
G^{(Q)}_{JT,abc}n_3^a
&\quad&\supset
&\quad&
H^{(Q)}_{JT,33}\,n_3^b n_3^c,
\\
\text{Level 2}: \quad&
G^{(Q)}_{JT,abc}n_3^a n_2^c
&\quad&\supset
&\quad&
- H^{(Q)}_{JT,3}\,n_3^b,
\\
&
G^{(Q)}_{JT,abc}n_2^b n_2^c
&\quad&\supset
&\quad&
H^{(Q)}_{JT,2}\,n_2^a,
\\
\text{Level 3}: \quad&
G^{(Q)}_{JT,abc}n_3^a n_2^b n_2^c
&\quad&=
&\quad&
- H^{(Q)}_{JT,0}.
\end{alignedat}
\end{equation}

\paragraph{Tensor-tensor}

\begin{equation}
\begin{alignedat}{3}
\text{Level 0}: \quad&
G^{(Q)}_{TT,abcd}
&\quad&\supset
&\quad&
H^{(Q)}_{TT,2233}\,n_2^a n_2^b n_3^c n_3^d,
\\
&
G^{(Q)}_{TT,abcd}
&\quad&\supset
&\quad&
H^{(Q)}_{TT,2\delta3}
\left(
\delta^{ac}n_2^b n_3^d
+\delta^{ad}n_2^b n_3^c
+\delta^{bc}n_2^a n_3^d
+\delta^{bd}n_2^a n_3^c
\right),
\\
&
G^{(Q)}_{TT,abcd}
&\quad&\supset
&\quad&
H^{(Q)}_{TT,\delta\delta}
\left(
\delta^{ac}\delta^{bd}
+
\delta^{ad}\delta^{bc}
\right),
\\
\text{Level 1}: \quad&
G^{(Q)}_{TT,abcd}n_2^d
&\quad&\supset
&\quad&
- H^{(Q)}_{TT,223}\,n_2^a n_2^b n_3^c
-
H^{(Q)}_{TT,2\delta}
\left(
n_2^a\delta^{bc}+n_2^b\delta^{ac}
\right),
\\
&
G^{(Q)}_{TT,abcd}n_3^a
&\quad&\supset
&\quad&
H^{(Q)}_{TT,233}\,n_2^b n_3^c n_3^d
+
H^{(Q)}_{TT,\delta3}
\left(
\delta^{bc}n_3^d+\delta^{bd}n_3^c
\right),
\\
\text{Level 2}: \quad&
G^{(Q)}_{TT,abcd}n_3^a n_3^b
&\quad&\supset
&\quad&
- H^{(Q)}_{TT,33}\,n_3^c n_3^d,
\\
&
G^{(Q)}_{TT,abcd}n_2^c n_2^d
&\quad&\supset
&\quad&
- H^{(Q)}_{TT,22}\,n_2^a n_2^b,
\\
&
G^{(Q)}_{TT,abcd}n_3^a n_2^d
&\quad&\supset
&\quad&
- H^{(Q)}_{TT,23}\,n_2^b n_3^c
-
H^{(Q)}_{TT,\delta}\,\delta^{bc},
\\
\text{Level 3}: \quad&
G^{(Q)}_{TT,abcd}n_3^a n_3^b n_2^d
&\quad&\supset
&\quad&
H^{(Q)}_{TT,3}\,n_3^c,
\\
&
G^{(Q)}_{TT,abcd}n_3^a n_2^c n_2^d
&\quad&\supset
&\quad&
- H^{(Q)}_{TT,2}\,n_2^b,
\\
\text{Level 4}: \quad&
G^{(Q)}_{TT,abcd}n_3^a n_3^b n_2^c n_2^d
&\quad&=
&\quad&
H^{(Q)}_{TT,0}.
\end{alignedat}
\end{equation}

There is a simple mnemonic for these definitions. To obtain a higher-level \(H\)-function, we contract only the indices of the operator inserted at point \(i\) with the corresponding vector \(n_i\). After this contraction, the correlator has the tensorial form of a correlator with an operator of lower spin. We then read off the coefficient using the same rule as in the lower-spin case.

For example, consider \(G^{(Q)}_{SJ,a}\). At level zero we read off the coefficient of the structure proportional to the vector associated with the scalar insertion at \(x_3\):
\begin{equation}
G^{(Q)}_{SJ,a}
\supset
H^{(Q)}_{SJ,3}\,n_3^a .
\end{equation}
At level one we contract the vector index with the unit vector associated with the insertion at \(x_2\). This contraction produces the relevant scalar structure:
\begin{equation}
G^{(Q)}_{SJ,a}n_2^a
=
H^{(Q)}_{SJ,0}.
\end{equation}
The same rule applies to the current-current correlator. At level one, contracting the index of the operator inserted at \(x_3\) leaves one free vector index:
\begin{equation}
G^{(Q)}_{JJ,ab}n_3^a .
\end{equation}
The resulting object has the same tensorial form as the scalar-current correlator, with the operator at \(x_3\) effectively reduced to a scalar insertion. We therefore read off the coefficient multiplying the vector associated with this effective scalar insertion:
\begin{equation}
-G^{(Q)}_{JJ,ab}n_3^a
\supset
H^{(Q)}_{JJ,3}\,n_3^b .
\end{equation}
Similarly, contracting the index of the operator inserted at \(x_2\) gives
\begin{equation}
G^{(Q)}_{JJ,ab} n_2^b
\supset
H^{(Q)}_{JJ,2}\,n_2^a .
\end{equation}
This rule fixes the choice of projections used above.

\section{Large-charge expansion}
\label{sec:large-charge-expansion}

\subsection{Leading order}
\label{subsec:leading-order}

We now discuss the expansion of the coefficient functions introduced in the previous section. In the \(s\)-channel, the four-point functions admit an expansion in powers of \(|z|\). This is the expansion in conformal primary multiplets exchanged between the two probe insertions. A primary operator of scaling dimension \(\Delta\) contributes with the leading power
\begin{equation}
|z|^{\Delta-\frac12(\Delta_{Q_{\rm in}}+\Delta_{Q_{\rm out}})} ,
\end{equation}
where \(Q_{\rm in}\) and \(Q_{\rm out}\) denote the charges of the incoming and outgoing heavy states. Descendants within the same conformal multiplet are then organized in additional powers of \(|z|\). The coefficient of the first descendant is fixed by the conformal algebra and is schematically of the form
\begin{equation}
\label{eq:DescendantGeneral}
\frac{
\Big[\Delta_{Q_{\rm in}}-\Delta+O(1)\Big]
\Big[\Delta_{Q_{\rm out}}-\Delta+O(1) \Big ]
}{
2\Delta
}.
\end{equation}
For each correlator, charge conservation fixes the charge of the exchanged states, denoted by \(Q_{\rm ex}\). The leading contribution in the corresponding \(s\)-channel is then given by the lowest-dimension operator in this charge sector, with scaling dimension \(\Delta_{Q_{\rm ex}}\). All other exchanged primaries are suppressed by powers of \(|z|^{\Delta-\Delta_{Q_{\rm ex}}}\), and therefore start contributing appreciably only at cylinder separations satisfying
\begin{equation}
|\tau| \sim  \f{1}{\Delta-\Delta_{Q_{\rm ex}}}
\end{equation}
At fixed nonzero \(\tau\), the \(s\)-channel expansion is consequently sensitive only to operators with
\begin{equation}
\Delta-\Delta_{Q_{\rm ex}}=O(1).
\end{equation}
In the macroscopic limit considered below,
\begin{equation}
|\tau|
\sim
\f{1}{\sqrt{|Q|}},
\end{equation}
the range of states that can contribute enlarges to
\begin{equation}
\Delta-\Delta_{Q_{\rm ex}}
=
O\!\left(\sqrt{|Q|}\right)
\end{equation}

Thus, the first descendant contribution of any such primary is suppressed in the large-charge expansion. Operators whose excitation energies grow faster than \(\sqrt{|Q|}\) are exponentially suppressed.

As a result, the leading contribution in the \(s\)-channel comes from the exchange of the lowest-dimension operator in the charge-\(Q_{\rm ex}\) sector. The leading-order coefficients involving the current and the stress tensor are fixed by the corresponding Ward identities. With the conventions introduced above, the only non-vanishing leading-order coefficient functions are
\begin{align}
\label{eq:LO_SS}
H_{SS}^{(Q),0}
&=
|\lambda_{q, Q}|^2\,
|z|^{\Delta_{Q+q}-\Delta_Q},
\\
\label{eq:LO_SJ}
H_{SJ,0}^{(Q),0}
&=
\lambda_{-q, Q+\frac{q}{2}}\,
\frac{Q+\frac q2}{\Omega_{d-1}}\,
|z|^{
\frac12\left(
\Delta_{Q+\frac q2}
-
\Delta_{Q-\frac q2}
\right)
},
\\
\label{eq:LO_JJ}
H_{JJ,0}^{(Q),0}
&=
\left( \frac{Q}{\Omega_{d-1}} \right)^2,
\\
\label{eq:LO_ST}
H_{ST,0}^{(Q),0}
&=
\lambda_{-q, Q+\frac{q}{2}}\,
\frac{\Delta_{Q+\frac q2}}{\Omega_{d-1}}\,
|z|^{
\frac12\left(
\Delta_{Q+\frac q2}
-
\Delta_{Q-\frac q2}
\right)
},
\\
\label{eq:LO_JT}
H_{JT,0}^{(Q),0}
&=
\frac{Q}{\Omega_{d-1}}\,
\frac{\Delta_Q}{\Omega_{d-1}},
\\
\label{eq:LO_TT}
H_{TT,0}^{(Q),0}
&=
\left( \frac{\Delta_Q}{\Omega_{d-1}} \right)^2 .
\end{align}
Here \(\Omega_{d-1}\) is the volume of the unit sphere \(S^{d-1}\), and we use the shorthand notation
\begin{equation}
\label{eqLambda_Definition}
\lambda_{q, Q} = \la Q +q| \phi_q | Q \ra.  
\end{equation}
All other leading-order coefficient functions vanish.

\subsection{Next-to-leading order \label{subsec:next-to-leading-order}}

At next-to-leading order there are two types of contributions. The first comes from the first descendant of the leading exchanged primary in the corresponding charge sector. The second comes from new primary states. The explicit descendant contribution in the 
\(F\)-basis is given in Appendix~\ref{app:F-descendants}. Applying the map from the \(F\)-basis to the \(H\)-basis gives the corresponding \(H^{(Q),1}\)-functions, collected in Appendix~\ref{app:H-descendants}.

For the new primary-state contribution, we restrict attention to \emph{parity-even} exchanged operators. We assume the same large-charge scaling as for the descendant of the leading exchanged primary. More precisely, in each channel the suppression is dictated by \eqref{eq:DescendantGeneral}. The exchanged primary states are traceless symmetric tensors \( {\cal O}_{a_1\dots a_\ell} \) with charge \(Q_{\rm ex}\).\footnote{Note that there are no mixed-symmetry tensors in $d=3$.} Their possible contributions are constrained by the three-point functions involving \(S\), \(J\), or \(T\), reviewed in Appendix~\ref{app:three-point-structures}. For \(\ell\geq 2\), there is no additional restriction on the dimension of the exchanged primary.

The low-spin cases require special treatment. For \(\ell=1\), conservation of the stress tensor implies that a spin-one primary can contribute to channels involving \(T\) only if
\begin{equation}
\Delta=\Delta_{Q_{\rm in}}+1.
\end{equation}
Such a primary is degenerate with the first descendant of the lowest-dimension charge-\(Q_{\rm in}\) primary. 

If we impose the stronger non-degeneracy assumption, which excludes degeneracy between the first descendant of the ground-state primary and the new spin-one primaries entering at this order, such a primary is excluded.

For \(\ell=0\), conservation of the current and of the stress tensor implies that channels involving \(J\) or \(T\) are nonzero only if
\begin{equation}
\Delta=\Delta_{Q_{\rm in}}.
\end{equation}
The minimal assumption that the lowest-dimension state in each large-charge sector is unique therefore excludes any additional spin-zero primary with this dimension. Hence spin-zero primary states do not contribute to channels involving \(J\) or \(T\).

It is useful to factor out the overall suppression dictated by the descendant contribution and to define normalized next-to-leading functions by
\begin{equation}
\label{eq:hDefinitionNLO}
\kappa_{SS}
=
\alpha\,\frac{3q^2}{8\sqrt {|Q|}},
\quad
\kappa_{SJ}
=
\frac{q}{2Q},
\quad
\kappa_{JJ}
=
\frac{2}{3\alpha |Q|^{3/2}},
\quad
\kappa_{ST}
=
\frac{3q}{4Q},
\quad
\kappa_{JT}
=
\frac{1}{\alpha |Q|^{3/2}},
\quad
\kappa_{TT}
=
\frac{3}{2\alpha |Q|^{3/2}}.
\end{equation}
The superscript \((Q)\) on \(h_{AB,r}\) has been dropped because the explicit \(Q\)-dependence of the next-to-leading contribution has been factored out into \(H^{(Q,0)}_{AB,0}\) and \(\kappa_{AB}\). Thus \(h_{AB,r}\) is kept fixed in the large-\(Q\) limit.

With these conventions, we introduce the notation
\begin{equation}
J_\ell^2=\frac{\ell(\ell+1)}{2},
\end{equation}
and
\begin{equation}
\omega_{\ell,i}=\Delta_{{\cal O}_{\ell,i}}-\Delta_{Q_{\rm ex}},
\end{equation}
where the label \(i\) runs over the Regge trajectories contributing at this order. Accordingly, both the OPE coefficients and the excitation energies depend on the spin \(\ell\) and on the trajectory label \(i\). The \(h\)-functions are then given by the following expressions.\footnote{The form of the spectral sums follows from the decomposition of charge-\(Q_{\rm ex}\) traceless symmetric primaries on the sphere and from the three-point tensor structures reviewed in Appendix~\ref{app:spectral-decomposition}.}

\paragraph{Scalar-scalar}

\begin{equation}
h_{SS}
=
3e^\tau\cos\theta
+
\sum_i \sum_{\ell=0}^\infty
(2\ell+1)|\lambda_{S,\ell,i}|^2
e^{\omega_{\ell,i}\tau}
C_\ell^{(1/2)}(\cos\theta).
\end{equation}

\paragraph{Scalar-current}

\begin{align}
\text{Level 0}: \quad
h_{SJ,3}
&=
\frac{3}{2}e^\tau
+
\sum_i\sum_{\ell=1}^\infty
(2\ell+1)\lambda_{S,\ell,i}^*\lambda_{J,\ell,i}
\frac{\omega_{\ell,i}}{2J_\ell^2}
e^{\omega_{\ell,i}\tau}
C_{\ell-1}^{(3/2)}(\cos\theta),
\\
\text{Level 1}: \quad
h_{SJ,0}
&=
3e^\tau\cos\theta
+
\sum_i\sum_{\ell=1}^\infty
(2\ell+1)\lambda_{S,\ell,i}^*\lambda_{J,\ell,i}
e^{\omega_{\ell,i}\tau}
C_\ell^{(1/2)}(\cos\theta).
\end{align}

\paragraph{Current-current}

\begin{align}
\text{Level 0}: \quad
h_{JJ,23}
&=
\sum_i\sum_{\ell=2}^{\infty}
(2\ell+1)|\lambda_{J,\ell,i}|^2
\frac{3\omega_{\ell,i}^2}{4J_\ell^4}
e^{\omega_{\ell,i}\tau}
C_{\ell-2}^{(5/2)}(\cos\theta),
\\
h_{JJ,\d}
&=
\frac{3}{4}e^\tau
+
\sum_i\sum_{\ell=1}^{\infty}
(2\ell+1)|\lambda_{J,\ell,i}|^2
\frac{\omega_{\ell,i}^2}{4J_\ell^4}
e^{\omega_{\ell,i}\tau}
C_{\ell-1}^{(3/2)}(\cos\theta),
\\
\text{Level 1}: \quad
h_{JJ,3}
&=
\frac{3}{2}e^\tau
+
\sum_i\sum_{\ell=1}^{\infty}
(2\ell+1)|\lambda_{J,\ell,i}|^2
\frac{\omega_{\ell,i}}{2J_\ell^2}
e^{\omega_{\ell,i}\tau}
C_{\ell-1}^{(3/2)}(\cos\theta),
\\
h_{JJ,2}
&=
\frac{3}{2}e^\tau
+
\sum_i\sum_{\ell=1}^{\infty}
(2\ell+1)|\lambda_{J,\ell,i}|^2
\frac{\omega_{\ell,i}}{2J_\ell^2}
e^{\omega_{\ell,i}\tau}
C_{\ell-1}^{(3/2)}(\cos\theta),
\\
\text{Level 2}: \quad
h_{JJ,0}
&=
3e^\tau\cos\theta
+
\sum_i\sum_{\ell=1}^{\infty}
(2\ell+1)|\lambda_{J,\ell,i}|^2
e^{\omega_{\ell,i}\tau}
C_\ell^{(1/2)}(\cos\theta).
\end{align}

\paragraph{Scalar-tensor}

\begin{align}
\text{Level 0}: \quad
h_{ST,33}
&=
\sum_i\sum_{\ell=2}^{\infty}
(2\ell+1)\lambda_{S,\ell,i}^*\lambda_{T,\ell,i}
\frac{3\left(J_\ell^2-\omega_{\ell,i}^2\right)}
{2J_\ell^2\left(J_\ell^2-1\right)}
e^{\omega_{\ell,i}\tau}
C_{\ell-2}^{(5/2)}(\cos\theta),
\\
\text{Level 1}: \quad
h_{ST,3}
&=
\frac{3}{2}
\left(
1+\sum_i\lambda_{S,1,i}^*\lambda_{T,1,i}
\right)e^\tau
+
\sum_i\sum_{\ell=2}^{\infty}
(2\ell+1)\lambda_{S,\ell,i}^*\lambda_{T,\ell,i}
\frac{\omega_{\ell,i}}{2J_\ell^2}
e^{\omega_{\ell,i}\tau}
C_{\ell-1}^{(3/2)}(\cos\theta),
\\
\text{Level 2}: \quad
h_{ST,0}
&=
3
\left(
1+\sum_i\lambda_{S,1,i}^*\lambda_{T,1,i}
\right)e^\tau\cos\theta
+
\sum_i\sum_{\ell=2}^{\infty}
(2\ell+1)\lambda_{S,\ell,i}^*\lambda_{T,\ell,i}
e^{\omega_{\ell,i}\tau}
C_\ell^{(1/2)}(\cos\theta).
\end{align}

\paragraph{Current-tensor}

\begin{align}
\text{Level 0}: \quad
h_{JT,233}
&=
\sum_i\sum_{\ell=3}^{\infty}
(2\ell+1)\lambda_{J,\ell,i}^*\lambda_{T,\ell,i}
\frac{15\omega_{\ell,i}\left(J_\ell^2-\omega_{\ell,i}^2\right)}
{4J_\ell^4\left(J_\ell^2-1\right)}
e^{\omega_{\ell,i}\tau}
C_{\ell-3}^{(7/2)}(\cos\theta),
\\
h_{JT,\d3}
&=
\sum_i\sum_{\ell=2}^{\infty}
(2\ell+1)\lambda_{J,\ell,i}^*\lambda_{T,\ell,i}
\frac{3\omega_{\ell,i}\left(J_\ell^2-\omega_{\ell,i}^2\right)}
{4J_\ell^4\left(J_\ell^2-1\right)}
e^{\omega_{\ell,i}\tau}
C_{\ell-2}^{(5/2)}(\cos\theta).
\end{align}

\begin{align}
\text{Level 1}: \quad
h_{JT,23}
&=
\sum_i\sum_{\ell=2}^{\infty}
(2\ell+1)\lambda_{J,\ell,i}^*\lambda_{T,\ell,i}
\frac{3\omega_{\ell,i}^2}{4J_\ell^4}
e^{\omega_{\ell,i}\tau}
C_{\ell-2}^{(5/2)}(\cos\theta),
\\
h_{JT,\d}
&=
\frac{3}{4}
\left(
1+\sum_i\lambda_{J,1,i}^*\lambda_{T,1,i}
\right)e^\tau
+
\sum_i\sum_{\ell=2}^{\infty}
(2\ell+1)\lambda_{J,\ell,i}^*\lambda_{T,\ell,i}
\frac{\omega_{\ell,i}^2}{4J_\ell^4}
e^{\omega_{\ell,i}\tau}
C_{\ell-1}^{(3/2)}(\cos\theta),
\\
h_{JT,33}
&=
\sum_i\sum_{\ell=2}^{\infty}
(2\ell+1)\lambda_{J,\ell,i}^*\lambda_{T,\ell,i}
\frac{3\left(J_\ell^2-\omega_{\ell,i}^2\right)}
{2J_\ell^2\left(J_\ell^2-1\right)}
e^{\omega_{\ell,i}\tau}
C_{\ell-2}^{(5/2)}(\cos\theta).
\end{align}

\begin{align}
\text{Level 2}: \quad
h_{JT,3}
&=
\frac{3}{2}
\left(
1+\sum_i\lambda_{J,1,i}^*\lambda_{T,1,i}
\right)e^\tau
+
\sum_i\sum_{\ell=2}^{\infty}
(2\ell+1)\lambda_{J,\ell,i}^*\lambda_{T,\ell,i}
\frac{\omega_{\ell,i}}{2J_\ell^2}
e^{\omega_{\ell,i}\tau}
C_{\ell-1}^{(3/2)}(\cos\theta),
\\
h_{JT,2}
&=
\frac{3}{2}
\left(
1+\sum_i\lambda_{J,1,i}^*\lambda_{T,1,i}
\right)e^\tau
+
\sum_i\sum_{\ell=2}^{\infty}
(2\ell+1)\lambda_{J,\ell,i}^*\lambda_{T,\ell,i}
\frac{\omega_{\ell,i}}{2J_\ell^2}
e^{\omega_{\ell,i}\tau}
C_{\ell-1}^{(3/2)}(\cos\theta),
\\
\text{Level 3}: \quad
h_{JT,0}
&=
3
\left(
1+\sum_i\lambda_{J,1,i}^*\lambda_{T,1,i}
\right)e^\tau\cos\theta
+
\sum_i\sum_{\ell=2}^{\infty}
(2\ell+1)\lambda_{J,\ell,i}^*\lambda_{T,\ell,i}
e^{\omega_{\ell,i}\tau}
C_\ell^{(1/2)}(\cos\theta).
\end{align}

\paragraph{Tensor-tensor}

\begin{align}
\text{Level 0}: \quad
h_{TT,2233}
&=
\sum_i\sum_{\ell=4}^{\infty}
(2\ell+1)|\lambda_{T,\ell,i}|^2
\frac{105\left(J_\ell^2-\omega_{\ell,i}^2\right)^2}
{4J_\ell^4\left(J_\ell^2-1\right)^2}
e^{\omega_{\ell,i}\tau}
C_{\ell-4}^{(9/2)}(\cos\theta),
\\
h_{TT,2\d3}
&=
\sum_i\sum_{\ell=3}^{\infty}
(2\ell+1)|\lambda_{T,\ell,i}|^2
\frac{15\left(J_\ell^2-\omega_{\ell,i}^2\right)^2}
{J_\ell^4\left(J_\ell^2-1\right)^2}
e^{\omega_{\ell,i}\tau}
C_{\ell-3}^{(7/2)}(\cos\theta),
\\
h_{TT,\d\d}
&=
\sum_i\sum_{\ell=2}^{\infty}
(2\ell+1)|\lambda_{T,\ell,i}|^2
\frac{3\left(J_\ell^2-\omega_{\ell,i}^2\right)^2}
{2J_\ell^4\left(J_\ell^2-1\right)^2}
e^{\omega_{\ell,i}\tau}
C_{\ell-2}^{(5/2)}(\cos\theta).
\end{align}

\begin{align}
\text{Level 1}: \quad
h_{TT,223}
&=
\sum_i\sum_{\ell=3}^{\infty}
(2\ell+1)|\lambda_{T,\ell,i}|^2
\frac{15\omega_{\ell,i}\left(J_\ell^2-\omega_{\ell,i}^2\right)}
{4J_\ell^4\left(J_\ell^2-1\right)}
e^{\omega_{\ell,i}\tau}
C_{\ell-3}^{(7/2)}(\cos\theta),
\\
h_{TT,2\d}
&=
\sum_i\sum_{\ell=2}^{\infty}
(2\ell+1)|\lambda_{T,\ell,i}|^2
\frac{3\omega_{\ell,i}\left(J_\ell^2-\omega_{\ell,i}^2\right)}
{4J_\ell^4\left(J_\ell^2-1\right)}
e^{\omega_{\ell,i}\tau}
C_{\ell-2}^{(5/2)}(\cos\theta),
\\
h_{TT,233}
&=
\sum_i\sum_{\ell=3}^{\infty}
(2\ell+1)|\lambda_{T,\ell,i}|^2
\frac{15\omega_{\ell,i}\left(J_\ell^2-\omega_{\ell,i}^2\right)}
{4J_\ell^4\left(J_\ell^2-1\right)}
e^{\omega_{\ell,i}\tau}
C_{\ell-3}^{(7/2)}(\cos\theta),
\\
h_{TT,\d3}
&=
\sum_i\sum_{\ell=2}^{\infty}
(2\ell+1)|\lambda_{T,\ell,i}|^2
\frac{3\omega_{\ell,i}\left(J_\ell^2-\omega_{\ell,i}^2\right)}
{4J_\ell^4\left(J_\ell^2-1\right)}
e^{\omega_{\ell,i}\tau}
C_{\ell-2}^{(5/2)}(\cos\theta).
\end{align}

\begin{align}
\text{Level 2}: \quad
h_{TT,33}
&=
\sum_i\sum_{\ell=2}^{\infty}
(2\ell+1)|\lambda_{T,\ell,i}|^2
\frac{3\left(J_\ell^2-\omega_{\ell,i}^2\right)}
{2J_\ell^2\left(J_\ell^2-1\right)}
e^{\omega_{\ell,i}\tau}
C_{\ell-2}^{(5/2)}(\cos\theta),
\\
h_{TT,22}
&=
\sum_i\sum_{\ell=2}^{\infty}
(2\ell+1)|\lambda_{T,\ell,i}|^2
\frac{3\left(J_\ell^2-\omega_{\ell,i}^2\right)}
{2J_\ell^2\left(J_\ell^2-1\right)}
e^{\omega_{\ell,i}\tau}
C_{\ell-2}^{(5/2)}(\cos\theta),
\\
h_{TT,23}
&=
\sum_i\sum_{\ell=2}^{\infty}
(2\ell+1)|\lambda_{T,\ell,i}|^2
\frac{3\left(J_\ell^2-\omega_{\ell,i}^2\right)}
{2J_\ell^2\left(J_\ell^2-1\right)}
e^{\omega_{\ell,i}\tau}
C_{\ell-2}^{(5/2)}(\cos\theta),
\\
h_{TT,\d}
&=
\frac{3}{4}
\left(
1+\sum_i|\lambda_{T,1,i}|^2
\right)e^\tau
+
\sum_i\sum_{\ell=2}^{\infty}
(2\ell+1)|\lambda_{T,\ell,i}|^2
\frac{\omega_{\ell,i}^2}{4J_\ell^4}
e^{\omega_{\ell,i}\tau}
C_{\ell-1}^{(3/2)}(\cos\theta).
\end{align}

\begin{align}
\text{Level 3}: \quad
h_{TT,3}
&=
\frac{3}{2}
\left(
1+\sum_i|\lambda_{T,1,i}|^2
\right)e^\tau
+
\sum_i\sum_{\ell=2}^{\infty}
(2\ell+1)|\lambda_{T,\ell,i}|^2
\frac{\omega_{\ell,i}}{2J_\ell^2}
e^{\omega_{\ell,i}\tau}
C_{\ell-1}^{(3/2)}(\cos\theta),
\\
h_{TT,2}
&=
\frac{3}{2}
\left(
1+\sum_i|\lambda_{T,1,i}|^2
\right)e^\tau
+
\sum_i\sum_{\ell=2}^{\infty}
(2\ell+1)|\lambda_{T,\ell,i}|^2
\frac{\omega_{\ell,i}}{2J_\ell^2}
e^{\omega_{\ell,i}\tau}
C_{\ell-1}^{(3/2)}(\cos\theta),
\\
\text{Level 4}: \quad
h_{TT,0}
&=
3
\left(
1+\sum_i|\lambda_{T,1,i}|^2
\right)e^\tau\cos\theta
+
\sum_i\sum_{\ell=2}^{\infty}
(2\ell+1)|\lambda_{T,\ell,i}|^2
e^{\omega_{\ell,i}\tau}
C_\ell^{(1/2)}(\cos\theta).
\end{align}

In the isolated spin-one terms above, the sum runs over all spin-one primaries with excitation energy \(\omega_{1,i}=1\) that can couple to the corresponding pair of probes. Their contributions have the same dependence on \(\tau\) and \(\theta\) as the first descendant of the ground-state primary and therefore enter only through the displayed linear combinations of fusion coefficients. If the stronger non-degeneracy assumption, which excludes degeneracy between the first descendant of the ground-state primary and the new spin-one primaries entering at this order, is imposed, all such spin-one primary contributions are absent, and the formulas reduce to the descendant contributions alone.

\section{Bootstrap equations \label{sec:bootstrap-equations}}

\subsection{Leading and subleading orders}

The crossing equations follow from exchanging the first and fourth operators in the original four-point function. In cylinder variables this exchanges the \(s\)- and \(u\)-channels and sends
\begin{equation}
\tau\to-\tau .
\end{equation}
For the charge assignments used above, the same operation is implemented on the family of correlators by reversing the sign of the large-charge parameter,
\begin{equation}
Q\to -Q .
\end{equation}
The resulting crossing equations for the \(H\)-basis are listed in Appendix~\ref{app:CrossingH-functions}.

Using equations \eqref{eq:LO_SS}-\eqref{eq:LO_TT}, the crossing equations imply\footnote{
This equality follows from crossing only at leading order in the large-\(Q\) expansion. A stronger all-orders statement can be obtained if the theory is invariant under charge conjugation. 
In such a theory one expects that
\begin{equation}
\lambda_{q,Q} = \lambda_{-q,-Q}.
\end{equation}
Combining this result with the Hermitian conjugation of \eqref{eqLambda_Definition} we obtain
\begin{equation}
\lambda_{-q, Q+\frac{q}{2}} 
=
\lambda^*_{-q, -Q+\frac{q}{2}}.
\end{equation}
Thus, after choosing compatible phases for the charged states, one can impose the exact relation
\begin{equation}
\lambda_{-q, Q+\frac{q}{2}} 
=
\lambda_{-q, -Q+\frac{q}{2}}.
\end{equation}}
\begin{equation}
\lambda_{-q,Q+\frac q2}
=
\lambda_{-q,-Q+\frac q2}
\end{equation}
at leading order in the large-\(Q\) expansion. There are several points which are useful to keep in mind when deriving the crossing equations for the normalized functions \(h_{AB,r}\) from the crossing equations for the original \(H\)-functions.

First, constant terms at the order considered here do not lead to new crossing constraints. Such terms multiply the same tensor structures as the leading solution and can be absorbed into a redefinition of the leading fusion coefficients. Equivalently, they amount to a correction to the overall normalization already fixed at leading order. Therefore, in writing the next-to-leading crossing equations for the \(h\)-functions, we keep only the non-constant part which cannot be absorbed into the leading crossing solution.

Second, in the scalar-scalar sector the next-to-leading crossing equation receives a contribution not only from \(h_{SS}\), but also from the mismatch between the dimensions of the leading exchanged states. Indeed, the crossing equation
\begin{equation}
H_{SS}^{(-Q)}
(-\tau, \cos\theta)
=
H_{SS}^{(Q)}(\tau, \cos\theta)
\end{equation}
gives
\begin{equation}
e^{\tau(\Delta_{Q+q}-\Delta_Q)}
\left[
1+\kappa_{SS}\,h_{SS}(\tau,\cos\theta)
\right]
=
e^{\tau(\Delta_Q-\Delta_{Q-q})}
\left[
1+\kappa_{SS}\,h_{SS}(-\tau,\cos\theta)
\right].
\end{equation}
With our normalization, the order \(1/\sqrt{|Q|}\) equation therefore takes the form
\begin{equation}
\tau+h_{SS}(\tau,\cos\theta)
=
-\tau+h_{SS}(-\tau,\cos\theta).
\end{equation}
After the redefinition
\begin{equation}
\label{eq:hSSRedefinition}
h_{SS}(\tau,\cos\theta)+\tau
\longrightarrow
h_{SS}(\tau,\cos\theta),
\end{equation}
this becomes the standard evenness condition
\begin{equation}
h_{SS}(-\tau,\cos\theta)
=
h_{SS}(\tau,\cos\theta).
\end{equation}

Third, in the mixed channels involving \(J\) or \(T\), one has to be careful about how the Ward-identity normalization is evaluated after crossing. The crossed correlator is evaluated at \(1/z\), with \(|z|<1\), or equivalently at \(-\tau\) with \(\tau<0\). In this radial ordering, the current or stress-tensor insertion surrounds the appropriate heavy state after the charged scalar insertion has been crossed. As a result, the Ward-identity factors in the crossed channel are fixed to be
\begin{equation}
Q+\frac q2
\end{equation}
for the current and
\begin{equation}
\Delta_{Q+\frac q2}
\end{equation}
for the stress tensor, rather than the naively expected factors obtained by simply replacing \(Q\to -Q\) inside the local \(s\)-channel expression. This point is essential for obtaining the correct crossing equations for the mixed \(SJ\) and \(ST\) functions.

The complete set of crossing equations for the \(h\)-functions is provided in Appendix~\ref{app:Crossingh-functions}.

\subsection{Algebraic equations}

We now explain how the crossing equations are converted into algebraic bootstrap equations. The idea is to combine the crossing equation with the asymptotic behavior near
\begin{equation}
\tau=\t=0,
\end{equation}
which is the singularity associated with the \(t\)-channel and is the only singularity. 

As in the EFT description, we assume that the discontinuity of the derivatives of the \(h\)-functions across \(\tau=0\) contains only contact singularities. More precisely, in the limit \(\eps\to0\), the discontinuity is assumed to approach a distribution supported at the coincident point on \(S^{d-1}\). A direct demonstration of this behavior in the EFT is given in Appendix~\ref{app:EFTContactSingularity}.

Let \(h\) be a function which, for \(\tau<0\), admits the expansion
\begin{equation}
\label{eq:hGenericExpansion}
h(\tau,\cos\t)
=
\sum_\ell
(2\ell+d-2)\,A_\ell\,
e^{\omega_\ell\tau}
C_{\ell-k}^{(d/2-1+k)}(\cos\t),
\end{equation}
and satisfies the crossing equation
\begin{equation}
\label{eq:hParityGeneric}
h(-\tau,\cos\t)
=
s\,h(\tau,\cos\t),
\qquad
s=\pm1.
\end{equation}
We also assume that near \(\tau=\t=0\) it has the following singular behavior
\begin{equation}
\label{eq:hAsymptoticGeneric}
h(\tau,\cos\t)
\underset{\tau,\t\to0}{=}
\f{
B_\b\!\left(
\f{\tau}{\sqrt{\tau^2+\t^2}}
\right)
}{
(\tau^2+\t^2)^{\f{d+\b}{2}}
}
+
\text{less singular terms},
\end{equation}
where \(B_\b(x)\) is a regular function.

Consider the integral of the discontinuity
\begin{align}
\label{eq:ImDefinition}
I_m
&=
\lim_{\eps\to0}
\int_0^\pi
\mathcal D_{m,\eps}(\t)\,
C_{\ell-k}^{(d/2-1+k)}(\cos\t)
\sin^{d-2+2k}\t\,d\t,
\end{align}
with
\begin{equation}
\label{eq:hDerivativeDiscontinuity}
\mathcal D_{m,\eps}(\t)
\equiv
\partial_\tau^m h(\tau,\cos\t)\Big|_{\tau=-\eps}
-
\partial_\tau^m h(\tau,\cos\t)\Big|_{\tau=\eps}.
\end{equation}
On one hand, using the spectral expansion \eqref{eq:hGenericExpansion} and the orthogonality of Gegenbauer
polynomials, we obtain
\begin{equation}
\label{eq:CrossingSpectral}
I_m
=
(2\ell+d-2)
N_{d,\ell-k,\f{d}{2}-1+k}
\left[
1-s(-1)^m
\right]
A_\ell\,\omega_\ell^m,
\end{equation}
where
\begin{equation}
\label{eq:GegenbauerNorm}
N_{d,\ell,\a}
=
\int_0^\pi
\left[
C_\ell^{(\a)}(\cos\t)
\right]^2
\sin^{2\a}\t\,d\t
=
\f{
2^{1-2\a}\pi\Gamma(\ell+2\a)
}{
(\ell+\a)\ell!\Gamma^2(\a)
}.
\end{equation}
On the other hand, we can evaluate the same integral using the contact singularity
\begin{equation}
\label{eq:ContactDistributionGeneric}
\lim_{\eps\to0}
\mathcal D_{m,\eps}(\vec n)
=
\sum_{r=0}^{\f{m+\b+1}{2}}
a_{m,r}
\left(-\vec\nabla^2\right)^r
\delta_{S^{d-1}}(\vec n).
\end{equation}
The upper limit follows from dimensional analysis.\footnote{In the applications below, the parity of \(m\) and the value of \(\b\) are such that \(r_{\rm max}\) is an integer.} It is useful to regard the factor \(\sin^{d-2}\t\,d\t\) as the standard radial part of the measure on \(S^{d-1}\), and to include the remaining factor in the test function
\begin{equation}
\label{eq:TestFunctionGeneric}
\Phi_{\ell,k}(\vec n)
\equiv
\sin^{2k}\t\,
C_{\ell-k}^{(d/2-1+k)}(\cos\t).
\end{equation}
Using \eqref{eq:ContactDistributionGeneric} and integrating by parts
on the sphere gives
\begin{align}
\label{eq:ContactPairingGeneric}
I_m
&=
\sum_{r=0}^{\f{m+\b+1}{2}}
a_{m,r}
\left[
\left(-\vec\nabla^2\right)^r
\Phi_{\ell,k}(\vec n)
\right]_{\theta=0}.
\end{align}
The small-\(\t\) expansion of the test function follows from the Gegenbauer differential equation:
\begin{equation}
\label{eq:GegenbauerSmallTheta}
\Phi_{\ell,k}(\t)
=
C_{\ell-k}^{(d/2-1+k)}(1)\,
\sum_{p=0}^{\infty}
\t^{2k+2p}
P_p\!\left(J_{d,\ell}^2\right),
\end{equation}
where \(P_p\) denotes a polynomial of degree \(p\), which may change
from line to line, and
\begin{equation}
\label{eq:JdDefinition}
J_{d,\ell}^2
\equiv
\f{\ell(\ell+d-2)}{d-1}.
\end{equation}
As a result, we obtain
\begin{equation}
\label{eq:CrossingAsymptotic}
I_m
=
\left[
1-s(-1)^m
\right]C_{\ell-k}^{(d/2-1+k)}(1)\,
P_{\f{m+\b+1}{2}-k}
\!\left(J_{d,\ell}^2\right).
\end{equation}
Comparing \eqref{eq:CrossingAsymptotic} with the spectral result \eqref{eq:CrossingSpectral}, absorbing all normalization
factors into the coefficients of the polynomial and separating the \(h\)-functions according to their parity under \(\tau\to-\tau\), taking
\begin{equation}
m=2n-1,
\end{equation}
and
\begin{equation}
m=2(n-1),
\end{equation}
for even and odd functions, respectively, we arrive to the following algebraic crossing equations.

For even functions,
\begin{equation}
\label{eq:BootstrapEvenGeneral}
A_\ell\,\omega_\ell^{2n-1}
=
P_{n+\f{\b}{2}-k}
\!\left(J_{d,\ell}^2\right),
\qquad
n\geq1.
\end{equation}
For odd functions,
\begin{equation}
\label{eq:BootstrapOddGeneral}
A_\ell\,\omega_\ell^{2n-2}
=
P_{n+\f{\b-1}{2}-k}
\!\left(J_{d,\ell}^2\right),
\qquad
n\geq1.
\end{equation}
When the index of the polynomial on the right-hand side is negative, the corresponding finite contribution is absent and the polynomial is
identically zero.

\subsection{Macroscopic limit \label{subsec:macroscopic-limit}}

As we saw above, in order to write the crossing equation for a function \(h_{AB}(\tau, \cos \theta)\), we need to know its asymptotic behavior near \(\tau=\t=0\). This asymptotic behavior is fixed by demanding that the correlators exist on the cylinder in the macroscopic limit
\begin{equation}
Q\to \infty,
\qquad
R\to \infty,
\end{equation}
with the energy density and the charge density kept fixed. Equivalently,
\begin{equation}
\frac{\Delta_Q}{Q^{d/(d-1)}} = \text{fixed}.
\end{equation}
In terms of cylinder variables, this is equivalent to demanding that the functions \(H\) have a finite limit as
\begin{equation}
\Delta_Q\to \infty,
\qquad
\tau Q^{1/(d-1)}=\text{fixed},
\qquad
\t Q^{1/(d-1)}=\text{fixed}.
\end{equation}
Therefore, using the scaling of the subleading contribution, and demanding that the four-point function \(G_{AB}\) has a well-defined macroscopic limit, we obtain an upper bound on how singular the correction to \(G_{AB}\) can be as \(\tau,\t\to 0\). The singularity can be softer, but below we assume saturation and study the resulting bootstrap equations.

The asymptotic behavior of the \(h\)-functions can be found as follows. For \(\t\to 0\),
\begin{equation}
n_3^a = n_2^a+\eta^a,
\qquad
\eta^a=O(\t).
\end{equation}
Expanding the correction to \(G_{AB}\) in powers of \(\eta^a\), one obtains different tensor structures. Near coincident insertions, each power of \(\eta^a\) can increase the singularity of the corresponding coefficient function by at most one power of the inverse distance. More precisely, the coefficient of a term containing \(p\) powers of \(\eta^a\) can be more singular than the coefficient of the term with no explicit powers of \(\eta\) in the same sector by at most
\begin{equation}
\frac{1}{(\tau^2+\theta^2)^{p/2}} .
\end{equation}
Thus, in the notation of \eqref{eq:hAsymptoticGeneric}, the value of \(\beta\) is determined by the singularity of the coefficient multiplying the tensor structure with no explicit powers of \(\eta^a\), and the number of explicit powers of \(\eta^a\) in the tensor structure under consideration.\footnote{As a check, the same asymptotic behavior is obtained from the large-\(\ell\) behavior of the spectral sums at \(\tau=0\), which determines the small-\(\theta\) singularity.}
The values of \(\beta\) for the functions entering the crossing equations are as follows.

For even functions,
\begin{align}
\beta=-2:\qquad
&
h_{SS},
\\[2mm]
\beta=0:\qquad
&
h_{SJ,3},\quad
h_{JJ,0},\quad
h_{ST,3},\quad
h_{JT,0},\quad
h_{TT,0},
\nonumber\\
&
h_{JJ,\d},\quad
h_{JT,\d},\quad
h_{TT,\d},\quad
h_{TT,\d\d},
\\[2mm]
\beta=2:\qquad
&
h_{JJ,23},\quad
h_{JT,23},\quad
h_{JT,33},
\nonumber\\
&
h_{TT,33},\quad
h_{TT,22},\quad
h_{TT,23},\quad
h_{TT,2\d3},
\\[2mm]
\beta=4:\qquad
&
h_{TT,2233}.
\end{align}

For odd functions,
\begin{align}
\beta=-1:\qquad
&
h_{SJ,0},\quad
h_{ST,0},
\\[2mm]
\beta=1:\qquad
&
h_{ST,33},\quad
h_{JJ,3},\quad
h_{JJ,2},\quad
h_{JT,3},\quad
h_{JT,2},
\nonumber\\
&
h_{TT,3},\quad
h_{TT,2},\quad
h_{JT,\d3},\quad
h_{TT,2\d},\quad
h_{TT,\d3},
\\[2mm]
\beta=3:\qquad
&
h_{JT,233},\quad
h_{TT,223},\quad
h_{TT,233}.
\end{align}

\subsection{Independent crossing equations \label{subsec:independent-crossing-equations}}

The crossing procedure described above produces one algebraic equation for each \(h\)-function. The full set is given in Appendix~\ref{app:full-crossing-equations}. Some of the equations differ only by an overall numerical coefficient multiplying the same function of \(\omega_{\ell,i}\) and \(J_\ell^2\). Such constants can be absorbed into the definition of the polynomial on the right-hand side. When two equations have the same structure, we keep the one that is valid on the larger spin domain. For instance, the equations coming from \(h_{JT,23}\) and \(h_{JT,\delta}\) have the same dependence on \(\omega_{\ell,i}\) and \(J_\ell^2\), but the latter also supplies the \(\ell=1\) constraint. We therefore keep the \(h_{JT,\delta}\) equation. For the subsequent analysis it is useful to work with a smaller independent set.

An independent set of crossing equations is as follows.

\paragraph{Scalar-scalar}

\begin{equation}
\delta_{\ell,0}\delta_{n,1}
+
\delta_{\ell,1}
+
\sum_i
|\lambda_{S,\ell,i}|^2\,
\omega_{\ell,i}^{2n-1}
=
P^{(SS)}_{n-1}(J_\ell^2),
\qquad
\ell\geq0.
\end{equation}

\paragraph{Scalar-current}

\begin{align}
\delta_{\ell,1}
+
\sum_i
\lambda_{S,\ell,i}^*\lambda_{J,\ell,i}
\frac{\omega_{\ell,i}^{2n}}{J_\ell^2}
&=
P^{(SJ,3)}_{n-1}(J_\ell^2),
\qquad
\ell\geq1,
\\
\delta_{\ell,1}
+
\sum_i
\lambda_{S,\ell,i}^*\lambda_{J,\ell,i}
\omega_{\ell,i}^{2n-2}
&=
P^{(SJ,0)}_{n-1}(J_\ell^2),
\qquad
\ell\geq1.
\end{align}

\paragraph{Current-current}

\begin{align}
\delta_{\ell,1}
+
\sum_i
|\lambda_{J,\ell,i}|^2
\frac{\omega_{\ell,i}^{2n+1}}{J_\ell^4}
&=
P^{(JJ,\delta)}_{n-1}(J_\ell^2),
\qquad
\ell\geq1,
\\
\delta_{\ell,1}
+
\sum_i
|\lambda_{J,\ell,i}|^2
\frac{\omega_{\ell,i}^{2n-1}}{J_\ell^2}
&=
P^{(JJ,3)}_{n-1}(J_\ell^2),
\qquad
\ell\geq1,
\\
\delta_{\ell,1}
+
\sum_i
|\lambda_{J,\ell,i}|^2
\omega_{\ell,i}^{2n-1}
&=
P^{(JJ,0)}_{n}(J_\ell^2),
\qquad
\ell\geq1.
\end{align}

\paragraph{Scalar-tensor}

For \(\ell\geq2\),
\begin{align}
\sum_i
\lambda_{S,\ell,i}^*\lambda_{T,\ell,i}
\frac{J_\ell^2-\omega_{\ell,i}^2}
{J_\ell^2\left(J_\ell^2-1\right)}
\omega_{\ell,i}^{2n-2}
&=
P^{(ST,33)}_{n-2}(J_\ell^2),
\\
\sum_i
\lambda_{S,\ell,i}^*\lambda_{T,\ell,i}
\frac{\omega_{\ell,i}^{2n}}{J_\ell^2}
&=
P^{(ST,3)}_{n-1}(J_\ell^2),
\\
\sum_i
\lambda_{S,\ell,i}^*\lambda_{T,\ell,i}
\omega_{\ell,i}^{2n-2}
&=
P^{(ST,0)}_{n-1}(J_\ell^2).
\end{align}

For \(\ell=1\),
\begin{equation}
P^{(ST,3)}_{n-1}(1)
=
P^{(ST,0)}_{n-1}(1)
=
1+\sum_i\lambda_{S,1,i}^*\lambda_{T,1,i}.
\end{equation}

\paragraph{Current-tensor}

For \(\ell\geq2\),
\begin{align}
\sum_i
\lambda_{J,\ell,i}^*\lambda_{T,\ell,i}
\frac{\omega_{\ell,i}^{2n+1}}{J_\ell^4}
&=
P^{(JT,\delta)}_{n-1}(J_\ell^2),
\\
\sum_i
\lambda_{J,\ell,i}^*\lambda_{T,\ell,i}
\frac{J_\ell^2-\omega_{\ell,i}^2}
{J_\ell^2\left(J_\ell^2-1\right)}
\omega_{\ell,i}^{2n-1}
&=
P^{(JT,33)}_{n-1}(J_\ell^2),
\\
\sum_i
\lambda_{J,\ell,i}^*\lambda_{T,\ell,i}
\frac{J_\ell^2-\omega_{\ell,i}^2}
{J_\ell^4\left(J_\ell^2-1\right)}
\omega_{\ell,i}^{2n-1}
&=
P^{(JT,\delta3)}_{n-2}(J_\ell^2),
\\
\sum_i
\lambda_{J,\ell,i}^*\lambda_{T,\ell,i}
\frac{\omega_{\ell,i}^{2n-1}}{J_\ell^2}
&=
P^{(JT,3)}_{n-1}(J_\ell^2),
\\
\sum_i
\lambda_{J,\ell,i}^*\lambda_{T,\ell,i}
\omega_{\ell,i}^{2n-1}
&=
P^{(JT,0)}_{n}(J_\ell^2).
\end{align}

For \(\ell=1\),
\begin{equation}
P^{(JT,\delta)}_{n-1}(1)
=
P^{(JT,3)}_{n-1}(1)
=
P^{(JT,0)}_{n}(1)
=
1+\sum_i\lambda_{J,1,i}^*\lambda_{T,1,i}.
\end{equation}

\paragraph{Tensor-tensor}

For \(\ell\geq2\),
\begin{align}
\sum_i
|\lambda_{T,\ell,i}|^2
\frac{\left(J_\ell^2-\omega_{\ell,i}^2\right)^2}
{J_\ell^4\left(J_\ell^2-1\right)^2}
\omega_{\ell,i}^{2n-1}
&=
P^{(TT,\delta\delta)}_{n-2}(J_\ell^2),
\\
\sum_i
|\lambda_{T,\ell,i}|^2
\frac{J_\ell^2-\omega_{\ell,i}^2}
{J_\ell^2\left(J_\ell^2-1\right)}
\omega_{\ell,i}^{2n-1}
&=
P^{(TT,33)}_{n-1}(J_\ell^2),
\\
\sum_i
|\lambda_{T,\ell,i}|^2
\frac{\omega_{\ell,i}^{2n+1}}{J_\ell^4}
&=
P^{(TT,\delta)}_{n-1}(J_\ell^2),
\\
\sum_i
|\lambda_{T,\ell,i}|^2
\omega_{\ell,i}^{2n-1}
&=
P^{(TT,0)}_{n}(J_\ell^2),
\\
\sum_i
|\lambda_{T,\ell,i}|^2
\frac{J_\ell^2-\omega_{\ell,i}^2}
{J_\ell^4\left(J_\ell^2-1\right)}
\omega_{\ell,i}^{2n-1}
&=
P^{(TT,2\delta)}_{n-2}(J_\ell^2),
\\
\sum_i
|\lambda_{T,\ell,i}|^2
\frac{\omega_{\ell,i}^{2n-1}}{J_\ell^2}
&=
P^{(TT,3)}_{n-1}(J_\ell^2).
\end{align}

For \(\ell=1\),
\begin{equation}
P^{(TT,\delta)}_{n-1}(1)
=
P^{(TT,0)}_{n}(1)
=
P^{(TT,3)}_{n-1}(1)
=
1+\sum_i|\lambda_{T,1,i}|^2.
\end{equation}
In the isolated spin-one equations above, the sums run over all spin-one primaries with excitation energy \(\omega_{1,i}=1\) that can couple to the corresponding pair of probes.

Whenever the index of a polynomial on the right-hand side is negative, the corresponding polynomial is identically zero. In particular, all equations involving \(P_{n-2}\) have a vanishing right-hand side at \(n=1\). There are no spin-zero primary contributions involving the current or the stress tensor.

\section{Solutions}
\label{sec:solutions}

\subsection{Minimal assumptions \label{sec:minimalAssumptions}}

As mentioned at the end of Section~\ref{subsec:next-to-leading-order}, for a fixed spin \(\ell\) there can be several charge-\(Q\) primary states contributing at the same order. For \(\ell\geq2\), we assume that the relevant charge-\(Q\) primary states organize into a finite number \(N\) of Regge trajectories. This assumption is motivated by analyticity in spin: away from the low-spin exceptional points, the CFT data can be continued as analytic functions of the spin~\cite{Caron-Huot:2017vep}. Thus, if the spectrum contains \(N\) analytic branches at large spin, we assume that the same \(N\) branches continue to every physical spin \(\ell\geq2\).

The cases \(\ell=0\) and \(\ell=1\) have to be treated separately. The spin-zero primary contribution appears only in the scalar-scalar channel. In channels involving \(T\), a spin-one primary can contribute only if its excitation energy is
\begin{equation}
\omega=1.
\end{equation}
We start with an immediate consequence of the tensor-tensor equations. Consider the equation coming from the \((TT,\delta\delta)\) structure. Its right-hand side is a polynomial of degree \(n-2\), while throughout the crossing equations
\begin{equation}
n=1,2,\ldots .
\end{equation}
For \(n=1\), this would formally give a polynomial of negative degree. This is an artifact of the derivation: it means that the corresponding singularity is not strong enough to contribute in the limit \(\eps\to0\). Therefore, the right-hand side vanishes at this order, and we obtain
\begin{equation}
\sum_{i=1}^{N}
|\lambda_{T,\ell,i}|^2
\frac{\omega_{\ell,i}
\left(J_\ell^2-\omega_{\ell,i}^2\right)^2}
{J_\ell^4\left(J_\ell^2-1\right)^2}
=
0,
\qquad
\ell\geq2.
\end{equation}
Every term in this sum is nonnegative. It follows that, for every \(i\),
\begin{equation}
|\lambda_{T,\ell,i}|^2
\left(J_\ell^2-\omega_{\ell,i}^2\right)^2
=
0.
\end{equation}
Consequently, every state contributing to the stress-tensor channel satisfies either
\begin{equation}
|\lambda_{T,\ell,i}|^2=0,
\end{equation}
or
\begin{equation}
\omega_{\ell,i}^2=J_\ell^2.
\end{equation}
The second option is precisely the spectrum of the standard Goldstone mode predicted by the large-charge conformal-superfluid effective theory. As a result, all equations proportional to \(J_\ell^2-\omega_{\ell,i}^2\) can be eliminated from the independent system.

For the remaining equations, it is convenient to introduce
\begin{equation}
x_i(w)
\equiv
\omega_{\ell,i}^2,
\qquad
w
\equiv
J_\ell^2.
\end{equation}
We also define
\begin{equation}
\label{eq:MomentsCoefficients}
S_i(w)
\equiv
|\lambda_{S,\ell,i}|^2\omega_{\ell,i},
\qquad
V_i(w)
\equiv
\frac{|\lambda_{J,\ell,i}|^2}{\omega_{\ell,i}},
\qquad
T_i(w)
\equiv
\frac{|\lambda_{T,\ell,i}|^2}{\omega_{\ell,i}}.
\end{equation}
Neglecting complex conjugation in the notation, we similarly write
\begin{equation}
\lambda_{S,\ell,i}\lambda_{J,\ell,i}
=
\sqrt{S_i(w)V_i(w)},
\qquad
\lambda_{S,\ell,i}\lambda_{T,\ell,i}
=
\sqrt{S_i(w)T_i(w)},
\qquad
\frac{\lambda_{J,\ell,i}\lambda_{T,\ell,i}}
{\omega_{\ell,i}}
=
\sqrt{V_i(w)T_i(w)}.
\end{equation}
The condition derived above becomes
\begin{equation}
T_i(w)\neq0
\quad\Longrightarrow\quad
x_i(w)=w.
\end{equation}
Thus, any trajectory that couples to the stress tensor must have the dispersion relation of the standard Goldstone mode of the conformal superfluid. The possibility that all trajectories satisfy \(x_i(w)\neq w\) is inconsistent. Indeed, in that case the condition above would imply
\begin{equation}
T_i(w)=0,
\qquad
i=1,\ldots,N,
\end{equation}
and hence
\begin{equation}
P_n^{(TT)}(w)=0
\end{equation}
for every physical value \(w\geq3\). Since \(P_n^{(TT)}(w)\) is a polynomial, it would then vanish identically and, in particular,
\begin{equation}
P_n^{(TT)}(1)=0.
\end{equation}
The spin-one crossing equation instead gives
\begin{equation}
P_n^{(TT)}(1)
=
1+\sum_i T_{1,i},
\qquad
n\geq1,
\end{equation}
where the sum runs over all spin-one primaries with excitation energy \(1\) that couple to the stress tensor. Since
\begin{equation}
T_{1,i}=|\lambda_{T,1,i}|^2\geq0,
\end{equation}
the right-hand side is strictly positive. We therefore obtain a contradiction. 

This establishes the result announced in the Introduction: under the minimal assumptions, the spectrum must contain at least one trajectory with the dispersion relation of the standard Goldstone mode of the conformal superfluid.

The different tensor structures within a fixed channel now lead to the same left-hand side, up to multiplication by powers of \(w\) and shifts of \(n\). This produces consistency conditions among the polynomials. To illustrate the reduction, consider the current-current sector. The three relevant equations have the form
\begin{align}
\sum_{i=1}^{N}
V_i(w)x_i^n(w)
&=
P_n^{(VV,0)}(w),
\\
\frac{1}{w}
\sum_{i=1}^{N}
V_i(w)x_i^n(w)
&=
P_{n-1}^{(VV,3)}(w),
\\
\frac{1}{w^2}
\sum_{i=1}^{N}
V_i(w)x_i^{n+1}(w)
&=
P_{n-1}^{(VV,\delta)}(w).
\end{align}
Multiplying the second equation by \(w\) gives
\begin{equation}
P_n^{(VV,0)}(w)
=
w\,P_{n-1}^{(VV,3)}(w),
\qquad
n\geq1.
\end{equation}
For the third equation, we first multiply by \(w^2\) and then shift \(n\to n-1\). This gives
\begin{equation}
P_n^{(VV,0)}(w)
=
w^2\,P_{n-2}^{(VV,\delta)}(w),
\qquad
n\geq2.
\end{equation}
Therefore,
\begin{equation}
P_n^{(VV,0)}(w)
=
w\,P_{n-1}^{(VV,3)}(w)
=
w^2\,P_{n-2}^{(VV,\delta)}(w),
\qquad
n\geq2.
\end{equation}

Doing the same in all channels gives the following six representative equations. Since the physical low-spin states need not coincide one-to-one with the analytically continued Regge trajectories, we display separately the equations for \(w\geq3\), \(w=1\), and \(w=0\).
We denote by \({\cal G}\) the set of trajectories whose dispersion relation coincides with that of the standard Goldstone mode of the conformal superfluid.\\
For \(w\geq3\),
\begin{align}
\sum_{i=1}^{N}
S_i(w)x_i^n(w)
&=
P_n^{(SS)}(w),
\qquad
n\geq0,
\\
\sum_{i=1}^{N}
\sqrt{S_i(w)V_i(w)}\,x_i^n(w)
&=
P_n^{(SV)}(w),
\qquad
n\geq0,
\\
\sum_{i=1}^{N}
V_i(w)x_i^n(w)
&=
P_n^{(VV)}(w),
\qquad
n\geq1,
\\
\sum_{i\in{\cal G}}
\sqrt{S_i(w)T_i(w)}\,w^n
&=
P_n^{(ST)}(w),
\qquad
n\geq0,
\\
\sum_{i\in{\cal G}}
\sqrt{V_i(w)T_i(w)}\,w^n
&=
P_n^{(VT)}(w),
\qquad
n\geq1,
\\
\sum_{i\in{\cal G}}
T_i(w)w^n
&=
P_n^{(TT)}(w),
\qquad
n\geq1.
\end{align}

For \(w=1\),
\begin{align}
1+
\sum_i
S_{1,i}x_{1,i}^n
&=
P_n^{(SS)}(1),
\qquad
n\geq0,
\\
1+
\sum_i
\sqrt{S_{1,i}V_{1,i}}\,x_{1,i}^n
&=
P_n^{(SV)}(1),
\qquad
n\geq0,
\\
1+
\sum_i
V_{1,i}x_{1,i}^n
&=
P_n^{(VV)}(1),
\qquad
n\geq1,
\\
1+
\sum_i
\sqrt{S_{1,i}T_{1,i}}
&=
P_n^{(ST)}(1),
\qquad
n\geq0,
\\
1+
\sum_i
\sqrt{V_{1,i}T_{1,i}}
&=
P_n^{(VT)}(1),
\qquad
n\geq1,
\\
1+
\sum_i
T_{1,i}
&=
P_n^{(TT)}(1),
\qquad
n\geq1.
\end{align}
Here \(i\) labels the physical spin-one primaries contributing at this order. In the channels involving \(T\), conservation requires \(x_{1,i}=1\).

For \(w=0\),
\begin{align}
\delta_{n,0}
+
\sum_i
S_{0,i}x_{0,i}^n
&=
P_n^{(SS)}(0),
\qquad
n\geq0,
\\
P_n^{(SV)}(0)
&=
P_n^{(ST)}(0)
=
0,
\qquad
n\geq0,
\\
P_n^{(VV)}(0)
&=
P_n^{(VT)}(0)
=
P_n^{(TT)}(0)
=
0,
\qquad
n\geq1.
\end{align}

The detailed relations among the structure polynomials will not be needed below. What matters is that, after choosing six representative polynomials, they obey the following consistency conditions:
\begin{equation}
\label{eq:ConsistencyPolynomials}
\renewcommand{\arraystretch}{1.45}
\begin{array}{c|c|c}
\text{Polynomial}
&
n=1
&
n\geq2
\\[4pt]
\hline
P_n^{(SS)}(w)
&
\text{none}
&
\text{none}
\\[4pt]
P_n^{(SV)}(w)
&
w\,p_{SV}
&
w\,\widetilde P_{n-1}^{(SV)}(w)
\\[4pt]
P_n^{(VV)}(w)
&
w\,p_{VV}
&
w^2\,\widetilde P_{n-2}^{(VV)}(w)
\\[4pt]
P_n^{(ST)}(w)
&
w\,p_{ST}
&
w\,\widetilde P_{n-1}^{(ST)}(w)
\\[4pt]
P_n^{(VT)}(w)
&
w\,p_{VT}
&
w^2\,\widetilde P_{n-2}^{(VT)}(w)
\\[4pt]
P_n^{(TT)}(w)
&
w\,p_{TT}
&
w^2\,\widetilde P_{n-2}^{(TT)}(w)
\end{array}
\end{equation}
In other words, the representative polynomials in all channels involving \(J\) or \(T\) are divisible by \(w\). In the \(VV\), \(VT\), and \(TT\) sectors, they are further divisible by \(w^2\) for \(n\geq2\).

Before moving forward, let us note one useful consequence of analyticity. Even though the number of physical states at \(w=0\) and \(w=1\) need not be equal to \(N\), their distinct energies are constrained by the analytic continuation of the polynomial data. For \(w\geq3\), the right-hand sides are polynomials in \(w\), and hence define analytic functions of \(w\). Consider, for example, an equation of the form
\begin{equation}
\sum_{i=1}^{N}
A_i(w)B_i(w)\,x_i^n(w)
=
P_n^{(AB)}(w).
\end{equation}
The corresponding generating function is\footnote{For the scalar-containing channels, the same argument applies even though the polynomials start from order zero. This changes only the numerator of the generating function and not the pole locations.}
\begin{equation}
\label{eq:GeneratingFunctionW}
W_{AB}(t,w)
\equiv
\sum_{n=1}^{\infty}
P_n^{(AB)}(w)t^n
=
\sum_{i=1}^{N}
\frac{
A_i(w)B_i(w)\,x_i(w)\,t
}{
1-tx_i(w)
}.
\end{equation}
Thus, for generic \(w\), the number of distinct poles in \(t\) is at most \(N\). Since the coefficients \(P_n^{(AB)}(w)\) are polynomials in \(w\), the same generating function can be analytically continued to \(w=0\) and \(w=1\). Therefore, after combining terms associated with the same pole and removing poles with vanishing total residue, the low-spin moment sequence cannot contain poles other than the analytically continued poles \(t=x_i(w)^{-1}\). Consequently, every distinct low-spin energy that contributes with nonzero total residue must be chosen from the analytically continued roots:
\begin{align}
x_{0,i}
&\in
\{x_1(0),\ldots,x_N(0)\},
\\
x_{1,i}
&\in
\{x_1(1),\ldots,x_N(1)\}.
\end{align}
This statement constrains the possible low-spin energies, but it does not determine the number of physical primaries at a given energy or their individual OPE coefficients. Degenerate primaries with the same energy contribute only through the corresponding total residue and cannot be distinguished by the moment equations alone. In particular, analyticity implies that there are at most \(N\) distinct contributing energies at each exceptional spin, but it does not imply that there are at most \(N\) physical primaries.

\subsection{Additional assumptions}

We have now established the main conclusion that follows from the minimal assumptions: at least one Regge trajectory must have the dispersion relation of the standard Goldstone mode of the conformal superfluid. We now impose the two additional assumptions stated in the Introduction, namely the condition in item~\ref{item:SpinZeroContinuation} and the non-degeneracy condition in item~\ref{item:NonDegeneracy}. These assumptions allow us to sharpen the solution and present the remaining crossing equations in a more transparent form.

We first use only the weaker part of the non-degeneracy assumption, namely non-degeneracy among primary states at fixed spin. Since every trajectory that couples to the stress tensor must satisfy
\begin{equation}
x_i(w)=w,
\end{equation}
two distinct such trajectories would produce two primary states with the same spin and excitation energy at every physical value \(w\geq3\). Non-degeneracy among primaries therefore implies that there is a unique Regge trajectory with the dispersion relation of the standard conformal-superfluid Goldstone. After relabelling, we denote it by \(i=1\):
\begin{equation}
x_1(w)=w.
\end{equation}
It is the only Regge trajectory that can couple to the stress tensor, and hence
\begin{equation}
T_i(w)=0,
\qquad
i=2,\ldots,N.
\end{equation}

This conclusion does not yet exclude spin-one primaries with excitation energy \(1\). Such primaries are degenerate with the first descendant of the ground-state primary rather than with another primary and are therefore allowed by the weaker form of the non-degeneracy assumption used above. We retain their contributions in the low-spin equations and impose the stronger non-degeneracy condition only later.

\subsubsection{\texorpdfstring{\(T\)}{T}-sector \label{subsec:T-sector}}

Let us first solve the part of the system involving the stress tensor. For \(w\geq3\), the \(TT\) equation becomes
\begin{equation}
T_1(w)\,w^n
=
P_n^{(TT)}(w).
\end{equation}
Using the consistency condition \eqref{eq:ConsistencyPolynomials}, we obtain
\begin{equation}
T_1(w)=p_{TT},
\end{equation}
and therefore
\begin{equation}
P_n^{(TT)}(w)=p_{TT}\,w^n.
\end{equation}
Similarly, the \(ST\) and \(VT\) equations give
\begin{equation}
S_1(w)=\frac{p_{ST}^2}{p_{TT}},
\qquad
V_1(w)=\frac{p_{VT}^2}{p_{TT}},
\end{equation}
or, equivalently,
\begin{equation}
P_{n-1}^{(ST)}(w)=p_{ST}\,w^{n-1},
\qquad
P_n^{(VT)}(w)=p_{VT}\,w^n.
\end{equation}

The \(w=1\) equations now give
\begin{align}
p_{ST}
&=
1+\sqrt{S_{1,1}T_{1,1}},
\\
p_{VT}
&=
1+\sqrt{V_{1,1}T_{1,1}},
\\
p_{TT}
&=
1+T_{1,1}.
\end{align}
Here we have used the weaker non-degeneracy assumption, according to which there can be at most one spin-one primary with excitation energy \(1\), and relabelled this possible state as \(i=1\). The tensor-sector polynomials therefore take the form
\begin{align}
P_{n-1}^{(ST)}(w)
&=
\left(
1+\sqrt{S_{1,1}T_{1,1}}
\right)
w^{n-1},
\\
P_n^{(VT)}(w)
&=
\left(
1+\sqrt{V_{1,1}T_{1,1}}
\right)
w^n,
\\
P_n^{(TT)}(w)
&=
\left(
1+T_{1,1}
\right)
w^n.
\end{align}
Correspondingly, the couplings of the unique conformal-superfluid Goldstone trajectory are
\begin{align}
T_1(w)
&=
1+T_{1,1},
\\
S_1(w)
&=
\frac{
\left(
1+\sqrt{S_{1,1}T_{1,1}}
\right)^2
}{
1+T_{1,1}
},
\\
V_1(w)
&=
\frac{
\left(
1+\sqrt{V_{1,1}T_{1,1}}
\right)^2
}{
1+T_{1,1}
}.
\end{align}

\subsubsection{\texorpdfstring{\(SV\)}{SV}-sector}
\label{subsec:SV-sector}

For \(w\geq3\), after isolating the unique trajectory that couples to \(T\), the equations in the \(SV\)-sector take the form
\begin{align}
S_1(w)\,w^{n-1}
+
\sum_{i=2}^{N}
S_i(w)\,x_i^{n-1}(w)
&=
P_{n-1}^{(SS)}(w),
\\
\sqrt{S_1(w)V_1(w)}\,w^{n-1}
+
\sum_{i=2}^{N}
\sqrt{S_i(w)V_i(w)}\,x_i^{n-1}(w)
&=
P_{n-1}^{(SV)}(w),
\\
V_1(w)\,w^n
+
\sum_{i=2}^{N}
V_i(w)\,x_i^n(w)
&=
P_n^{(VV)}(w),
\qquad
n\geq1.
\end{align}
Using the assumption in item~\ref{item:SpinZeroContinuation},
\begin{equation}
x_i(0)\neq0,
\qquad
i=2,\ldots,N,
\end{equation}
together with the \(VV\) polynomial consistency conditions \eqref{eq:ConsistencyPolynomials}, one can prove that the non-Goldstone trajectories do not couple to the current:
\begin{equation}
V_i(w)=0,
\qquad
i=2,\ldots,N.
\end{equation}
The proof is given in Appendix~\ref{app:SV-current-decoupling}. The argument is similar in spirit to the use of divisibility conditions in~\cite{Kiaee:2025jly}, where analogous polynomial constraints were used to show that the characteristic polynomial obeys \(Q_N(0)=0\).

It follows that the \(SV\) and \(VV\) polynomials receive contributions only from the trajectory of the standard Goldstone mode of the conformal superfluid:
\begin{align}
P_{n-1}^{(SV)}(w)
&=
\sqrt{S_1(w)V_1(w)}\,w^{n-1},
\\
P_n^{(VV)}(w)
&=
V_1(w)\,w^n.
\end{align}
Using the results of the \(T\)-sector, these become
\begin{align}
P_{n-1}^{(SV)}(w)
&=
\frac{p_{ST}p_{VT}}{p_{TT}}\,w^{n-1},
\\
P_n^{(VV)}(w)
&=
\frac{p_{VT}^2}{p_{TT}}\,w^n.
\end{align}

The energy of any physical spin-one primary contributing to these channels must be chosen from the analytically continued Regge trajectories. Since the non-Goldstone trajectories do not couple to the current, any spin-one primary with a nonzero current coupling must belong to the trajectory of the standard conformal-superfluid Goldstone and therefore has
\begin{equation}
x_{1,1}=x_1(1)=1.
\end{equation}
Moreover, non-degeneracy among primary states implies that there can be at most one such spin-one primary. By relabelling, we denote it by \(i=1\).

The \(w=1\) constraint in the \(VV\) channel is therefore
\begin{equation}
1+V_{1,1}
=
\frac{
\left(1+\sqrt{V_{1,1}T_{1,1}}\right)^2
}{
1+T_{1,1}
}.
\end{equation}
Equivalently,
\begin{equation}
V_{1,1}=T_{1,1}.
\end{equation}
It follows that
\begin{equation}
p_{VT}=p_{TT}=1+T_{1,1},
\end{equation}
and
\begin{equation}
V_1(w)=1+T_{1,1}.
\end{equation}
The \(w=1\) constraint in the \(SV\) channel is automatically satisfied. And as a result, 
\begin{align}
P_{n-1}^{(SV)}(w)
&=
P_{n-1}^{(ST)}(w)
=
\left(
1+\sqrt{S_{1,1}T_{1,1}}
\right)w^{n-1},
\\
P_n^{(VV)}(w)
&=
P_n^{(VT)}(w)
=
P_n^{(TT)}(w)
=
\left(
1+T_{1,1}
\right)w^n.
\end{align}

Thus, under the additional assumptions used so far, the non-Goldstone trajectories contribute only to the scalar-scalar channel at this order. The only remaining part of the system is the scalar-scalar equation,
\begin{equation}
P_n^{(SS)}(w)
=
S_1(w)w^n
+
\sum_{i=2}^{N}
S_i(w)x_i^n(w),
\qquad
n\geq0,
\end{equation}
where
\begin{equation}
S_1(w)
=
\frac{
\left(
1+\sqrt{S_{1,1}T_{1,1}}
\right)^2
}{
1+T_{1,1}
}.
\end{equation}
It is convenient to subtract the known conformal-superfluid Goldstone contribution and define
\begin{equation}
R_n(w)
\equiv
P_n^{(SS)}(w)
-
S_1(w)w^n
=
\sum_{i=2}^{N}
S_i(w)x_i^n(w).
\end{equation}
The remaining problem is the standard finite-trajectory recurrence problem, see~\cite{Jafferis:2017zna,Kiaee:2025jly}, now of order \(N-1\). Let
\begin{equation}
Q_k(w),
\qquad
k=1,\ldots,N-1,
\end{equation}
be the elementary symmetric polynomials in the non-Goldstone roots
\begin{equation}
x_2(w),\ldots,x_N(w).
\end{equation}
Equivalently, these roots are determined by
\begin{equation}
\label{eq:ResidualCharacteristicPolynomial}
x^{N-1}
-
Q_1(w)x^{N-2}
+
Q_2(w)x^{N-3}
+\cdots
+
(-1)^{N-1}Q_{N-1}(w)
=
0.
\end{equation}
The residual sequence \(R_n(w)\) satisfies the order-\(N-1\) recurrence relation
\begin{align}
\label{eq:SSResidualPolynomialRecurrence}
R_{n+N-1}(w)
&=
Q_1(w)R_{n+N-2}(w)
-
Q_2(w)R_{n+N-3}(w)
+\cdots
+
(-1)^NQ_{N-1}(w)R_n(w).
\end{align}
Once the roots \(x_2,\ldots,x_N\) are known, the remaining scalar coefficients \(S_i(w)\), with \(i=2,\ldots,N\), are fixed by the first \(N-1\) residual polynomials:
\begin{equation}
\begin{pmatrix}
1 & 1 & \cdots & 1 \\
x_2 & x_3 & \cdots & x_N \\
\vdots & \vdots & \ddots & \vdots \\
x_2^{N-2} & x_3^{N-2} & \cdots & x_N^{N-2}
\end{pmatrix}
\begin{pmatrix}
S_2 \\
S_3 \\
\vdots \\
S_N
\end{pmatrix}
=
\begin{pmatrix}
R_0 \\
R_1 \\
\vdots \\
R_{N-2}
\end{pmatrix}.
\end{equation}

It remains to examine the low-spin scalar constraints. Let \(N_1\) denote the number of analytically continued trajectories satisfying\footnote{The number \(N_1\) counts analytically continued Regge trajectories and not independent physical spin-one primaries with excitation energy \(\omega=1\). By non-degeneracy among primaries, there can be at most one such physical primary, which we label by \(i=1\).}
\begin{equation}
x_i(1)=1,
\qquad
i=1,\ldots,N_1,
\end{equation}
where \(i=1\) denotes the trajectory of the standard Goldstone mode of the conformal superfluid.

The \(SS\) equation at \(w=1\) gives
\begin{equation}
1+S_{1,1}
+
\sum_{i=2}^{N}
S_{1,i}\,x_{1,i}^{n-1}
=
\frac{
\left(1+\sqrt{S_{1,1}V_{1,1}}\right)^2
}{
1+V_{1,1}
}
+
\sum_{i=2}^{N}
S_i(1)\,x_i^{n-1}(1).
\end{equation}
Comparing the contributions associated with different values of \(x_i(1)\), and using the \(ST\) equation at \(w=1\), gives
\begin{equation}
\label{eq:Scalarw1x1Constraint}
\left(
\sqrt{S_{1,1}}-\sqrt{V_{1,1}}
\right)^2
=
\left(1+V_{1,1}\right)
\sum_{i=2}^{N_1}
S_i(1),
\end{equation}
together with
\begin{equation}
\label{eq:Spin1Remaining}
\sum_{i=N_1+1}^{N}
S_{1,i}\,x_{1,i}^{n-1}
=
\sum_{i=N_1+1}^{N}
S_i(1)\,x_i^{n-1}(1).
\end{equation}

The \(w=0\) constraint gives
\begin{equation}
\delta_{n,0}
+
\sum_{i=2}^{N}
S_{0,i}\,x_{0,i}^{n}
=
\delta_{n,0}
\frac{
\left(1+\sqrt{S_{1,1}V_{1,1}}\right)^2
}{
1+V_{1,1}
}
+
\sum_{i=2}^{N}
S_i(0)\,x_i^n(0).
\end{equation}
The uniqueness of the lowest-dimension state in the charge-\(Q\) sector excludes any additional physical spin-zero primary with zero excitation energy. Therefore, every physical spin-zero primary contributing to the sum has nonzero energy, and comparison of the nonzero-energy contributions gives
\begin{equation}
\label{eq:Spin0Remaining1}
\sum_{i=2}^{N}
S_{0,i}\,x_{0,i}^{n}
=
\sum_{i=2}^{N}
S_i(0)\,x_i^n(0).
\end{equation}
The zero-energy contribution gives
\begin{equation}
\left(
1+\sqrt{S_{1,1}V_{1,1}}
\right)^2
=
1+V_{1,1}.
\end{equation}

Combining this relation with \eqref{eq:Scalarw1x1Constraint}, we find two possibilities. If
\begin{equation}
S_{1,1}\geq V_{1,1},
\end{equation}
then
\begin{equation}
\label{eq:Spin0Remaining3}
S_{1,1}
=
\sum_{i=2}^{N_1}
S_i(1),
\qquad
V_{1,1}=0.
\end{equation}
If instead
\begin{equation}
S_{1,1}<V_{1,1},
\end{equation}
then
\begin{equation}
\label{eq:Spin0Remaining2}
S_{1,1}
=
\sum_{i=2}^{N_1}
S_i(1)
=
\frac{
V_{1,1}
}{
\left(
1+\sqrt{1+V_{1,1}}
\right)^2
}.
\end{equation}

All polynomials therefore take the form
\begin{align}
P_{n-1}^{(SS)}(w)
&=
w^{n-1}
+
\sum_{i=2}^{N}
S_i(w)\,x_i^{n-1}(w),
\\
P_{n-1}^{(SV)}(w)
&=
P_{n-1}^{(ST)}(w)
=
\sqrt{1+V_{1,1}}\,w^{n-1},
\\
P_n^{(VV)}(w)
&=
P_n^{(VT)}(w)
=
P_n^{(TT)}(w)
=
\left(1+V_{1,1}\right)w^n.
\end{align}

\paragraph{Stronger non-degeneracy}

If we impose the stronger form of the non-degeneracy assumption, which excludes degeneracy between the first descendant of the ground-state primary and the new spin-one primary entering at this order, the possible spin-one primary with excitation energy \(1\) is excluded. Consequently,
\begin{equation}
S_{1,1}
=
V_{1,1}
=
T_{1,1}
=
0.
\end{equation}
The polynomials then reduce to
\begin{align}
P_{n-1}^{(SS)}(w)
&=
w^{n-1}
+
\sum_{i=2}^{N}
S_i(w)\,x_i^{n-1}(w),
\\
P_{n-1}^{(SV)}(w)
&=
P_{n-1}^{(ST)}(w)
=
w^{n-1},
\\
P_n^{(VV)}(w)
&=
P_n^{(VT)}(w)
=
P_n^{(TT)}(w)
=
w^n.
\end{align}
Thus, under this assumption, every channel involving the current or the stress tensor is completely fixed by the trajectory of the standard Goldstone mode of the conformal superfluid. All remaining freedom is confined to the scalar-scalar channel, where the additional Regge trajectories may still contribute.

The remaining constraints \eqref{eq:Spin1Remaining} and \eqref{eq:Spin0Remaining1} do not lead to an illuminating general solution for arbitrary \(N\). To illustrate more explicitly how the remaining scalar channel freedom is organized, we now turn to the simplest concrete cases.

\subsubsection{One Regge trajectory}
\label{subsec:one-regge-trajectory}

For \(N=1\), there are no non-Goldstone trajectories. Hence \(N_1=1\), and the sum on the right-hand side of \eqref{eq:Scalarw1x1Constraint} is empty. The weak form of the non-degeneracy assumption then gives
\begin{equation}
S_{1,1}
=
V_{1,1}
=
T_{1,1}
=
0.
\end{equation}
Therefore, the unique solution is
\begin{align}
P_{n-1}^{(SS)}(w)
&=
P_{n-1}^{(SV)}(w)
=
P_{n-1}^{(ST)}(w)
=
w^{n-1},
\\
P_n^{(VV)}(w)
&=
P_n^{(VT)}(w)
=
P_n^{(TT)}(w)
=
w^n.
\end{align}
Thus, already under the weak form of non-degeneracy, all channels are completely fixed by the trajectory of the standard Goldstone mode of the conformal superfluid. The stronger non-degeneracy assumption does not impose any further restriction in this case.

\subsubsection{Two Regge trajectories}
\label{subsec:two-regge-trajectories}

For \(N=2\), the residual characteristic polynomial has degree one. Therefore,\footnote{By the assumption in item~\ref{item:SpinZeroContinuation}, \(m^2\neq0\).}
\begin{equation}
x_2(w)=c^2w+m^2.
\end{equation}
The residual sequence therefore has the form
\begin{equation}
R_{n-1}(w)
=
S_{0,2}
\left(c^2w+m^2\right)^{n-1},
\qquad
n\geq1.
\end{equation}
There are two cases, depending on whether the analytic continuation of the second trajectory reaches \(x=1\) at \(w=1\).

\paragraph{Case 1: \(c^2+m^2\neq1\) \ :}

In this case \(N_1=1\), and the weak form of non-degeneracy gives
\begin{equation}
S_{1,1}
=
V_{1,1}
=
T_{1,1}
=
0.
\end{equation}
The polynomials are therefore
\begin{align}
P_{n-1}^{(SS)}(w)
&=
w^{n-1}
+
S_{0,2}
\left(c^2w+m^2\right)^{n-1},
\\
P_{n-1}^{(SV)}(w)
&=
P_{n-1}^{(ST)}(w)
=
w^{n-1},
\\
P_n^{(VV)}(w)
&=
P_n^{(VT)}(w)
=
P_n^{(TT)}(w)
=
w^n.
\end{align}
The spin-zero and spin-one scalar constraints give
\begin{equation}
S_{1,2}=S_{0,2}.
\end{equation}
Thus, all channels involving the current or the stress tensor are again completely fixed by the trajectory of the standard Goldstone mode of the conformal superfluid, while the second trajectory contributes only to the scalar-scalar channel. The stronger non-degeneracy assumption does not modify this solution.

\paragraph{Case 2: \(c^2+m^2=1\) \ :}

In this case \(N_1=2\), so the second analytically continued trajectory also contributes to the \(x=1\) part of the spin-one scalar constraint. Under the weak form of non-degeneracy, there are two branches.

The first branch \eqref{eq:Spin0Remaining3} has
\begin{equation}
V_{1,1}
=
T_{1,1}
=
0,
\qquad
S_{1,1}
=
S_{0,2}.
\end{equation}
The polynomials are
\begin{align}
P_{n-1}^{(SS)}(w)
&=
w^{n-1}
+
S_{0,2}
\left(c^2w+m^2\right)^{n-1},
\\
P_{n-1}^{(SV)}(w)
&=
P_{n-1}^{(ST)}(w)
=
w^{n-1},
\\
P_n^{(VV)}(w)
&=
P_n^{(VT)}(w)
=
P_n^{(TT)}(w)
=
w^n.
\end{align}
The second branch \eqref{eq:Spin0Remaining2} has
\begin{equation}
S_{1,1}
=
S_{0,2}
=
\frac{
V_{1,1}
}{
\left(
1+\sqrt{1+V_{1,1}}
\right)^2
},
\qquad
T_{1,1}=V_{1,1},
\end{equation}
and the polynomials are
\begin{align}
P_{n-1}^{(SS)}(w)
&=
w^{n-1}
+
\frac{
V_{1,1}
}{
\left(
1+\sqrt{1+V_{1,1}}
\right)^2
}
\left(c^2w+m^2\right)^{n-1},
\\
P_{n-1}^{(SV)}(w)
&=
P_{n-1}^{(ST)}(w)
=
\sqrt{1+V_{1,1}}\,w^{n-1},
\\
P_n^{(VV)}(w)
&=
P_n^{(VT)}(w)
=
P_n^{(TT)}(w)
=
\left(1+V_{1,1}\right)w^n.
\end{align}
Purely scalar correlators alone do not reveal whether the normalization \(S_{0,2}\) is independent or tied to \(V_{1,1}\) through the second branch. Correlators involving spinning probes distinguish the two branches, since only the second branch modifies their normalization.

If the stronger form of non-degeneracy is imposed, the physical spin-one primary with excitation energy \(1\) is excluded:
\begin{equation}
S_{1,1}
=
V_{1,1}
=
T_{1,1}
=
0.
\end{equation}
Equation~\eqref{eq:Scalarw1x1Constraint} then also requires
\begin{equation}
S_2(1)=S_{0,2}=0.
\end{equation}
Hence, when \(c^2+m^2=1\), the second trajectory has vanishing spectral weight under the stronger non-degeneracy assumption, and the solution reduces effectively to the one-trajectory solution.

\section{Conclusions}
\label{sec:conclusions}

In this paper, we study the large-charge bootstrap equations in three-dimensional CFTs with a global \(U(1)\) symmetry. We consider four-point functions with two large-charge operators and two light probes chosen from a scalar operator, the conserved current \(J^a\), and the stress tensor \(T^{ab}\). This leads to a coupled system of bootstrap equations in the \(SS\), \(SJ\), \(JJ\), \(ST\), \(JT\), and \(TT\) channels. After writing the corresponding \(s\)-channel expansions to first subleading order in the large-charge expansion, we impose the smoothness conditions that follow from crossing at the boundary of the \(s\)-channel convergence region, together with the existence of the macroscopic limit on the cylinder.

Under the minimal assumptions stated in the Introduction, the tensor-tensor channel implies that every Regge trajectory that couples to the stress tensor must have the dispersion relation of the standard Goldstone mode of the conformal superfluid. Consistency of the low-spin equations then requires that at least one such trajectory be present. This provides a direct bootstrap derivation of the existence of the standard conformal-superfluid Goldstone trajectory.

The additional assumptions sharpen this result. Non-degeneracy among primary states implies that the conformal-superfluid Goldstone trajectory is unique. The assumption that no non-Goldstone trajectory reaches zero energy after analytic continuation to 
\(w=0\) then implies that all non-Goldstone trajectories decouple from the current and the stress tensor at this order. They can contribute only to the scalar-scalar channel. If the stronger form of non-degeneracy, which excludes degeneracy between the first descendant of the ground-state primary and the new spin-one primary entering at this order, is imposed, every channel involving the current or the stress tensor is completely fixed by the standard conformal-superfluid Goldstone trajectory. All remaining freedom is confined to the scalar-scalar channel.

The bootstrap equations therefore come close to establishing that, within the class of theories satisfying the assumptions of this paper and within the finite-energy sector considered here, the most general solution is the conformal-superfluid EFT accompanied by additional light fields. What remains unconstrained is precisely the scalar-channel data associated with those additional trajectories.

This conclusion should not be interpreted as saying that the conformal superfluid is the only possible large-charge phase. Fermionic theories may instead admit a large-charge sector described by a Fermi liquid on the sphere. Such a system contains an infinite number of low-energy Regge trajectories and therefore lies outside the finite-trajectory framework assumed in this paper.

Conformal solids provide another important class of examples outside the present analysis. In that case, parity-odd operators associated with the transverse phonon contribute to the relevant OPEs already at the order considered here. Since we restricted the exchanged operators in the \(s\)- and \(u\)-channel OPEs to the parity-even sector, the conformal-solid solution is not contained in the bootstrap system analyzed in this paper. Extending the analysis to include the parity-odd sector, and thereby incorporating conformal solids, is an important direction for future work.

Our analysis is also insensitive to states whose excitation energies are of order \(\sqrt{|Q|}\) or larger. The algebraic equations derived in this paper constrain the sector of states whose excitation energies above the lowest operator in the relevant charge sector remain finite in the large-charge limit,
\begin{equation}
\Delta-\Delta_{Q_{\rm ex}}=O(1).
\end{equation}
States with excitation energies of order \(\sqrt{|Q|}\) are, in principle, visible in the macroscopic limit, as explained in Section~\ref{subsec:leading-order}. However, their contribution to any four-point function is trivial in this limit. Even though the exponential factor
\begin{equation}
e^{-|\tau|(\Delta-\Delta_{Q_{\rm ex}})}
\end{equation}
remains finite, the overall contribution is suppressed by a power of \(|Q|\). Since only finitely many Regge trajectories are assumed to contribute at the order considered, the sum over trajectories cannot compensate for this suppression.

States with excitation energies of order \(\sqrt{|Q|}\) or higher therefore do not enter the algebraic equations derived here unless their spectral weights remain unsuppressed in the large-charge limit. Our conclusions concern the finite-energy sector and do not exclude the existence of additional scaling sectors with parametrically larger excitation energies.

\section*{Acknowledgements}

This work is supported by the National Science Foundation under Award No. 2310243.

\newpage
\appendix

\section{\texorpdfstring{\(F\)}{F}-basis \label{app:F-basis}}

In this appendix we present the tensor decomposition of the large-charge two-point functions in the \(F\)-basis. This basis is built directly from the vectors \(n_2^a\), \(n_3^a\), and the flat metric \(\delta^{ab}\). It is therefore the most direct basis from the point of view of tensor structures on the cylinder.  All coefficient functions \(F\) depend on the two conformal invariants, which we write as \((\tau,\cos\t)\). In correlators involving the stress tensor, we also use the symmetry and tracelessness of \(T^{ab}\). As a result, some tensor structures are related, and some coefficient functions are expressed in terms of the remaining ones.

\paragraph{Scalar-scalar}

\begin{equation}
G_{SS}(\tau, \cos \t) = F_{SS}(\tau, \cos \t).
\end{equation}

From now on, to simplify the notation, we will suppress the arguments of the coefficient functions. Thus all functions \(F\) below are understood to depend on \((\tau,\cos\t)\).

\paragraph{Scalar-current}

\begin{align}
G^a_{SJ} = n_3^a F_{SJ, 3}(\tau, \cos \t) + n_2^a F_{SJ, 2}(\tau, \cos \t)
\end{align}

\paragraph{Current-current}

\begin{align}
G^{ab}_{JJ} = n_3^a n_3^b F_{JJ, 33} + n_3^a n_2^b F_{JJ, 32} + n_2^a n_3^b F_{JJ,23} + n_2^a n_2^b F_{JJ, 22} + \d^{ab} F_{JJ, \d}.
\end{align}

\paragraph{Scalar-tensor}

\begin{align}
G^{ab}_{ST}  & = n_3^a n_3^b F_{ST, 33} + n_2^a n_2^b F_{ST, 22} + n_3^a n_2^b F_{ST, 32} + n_2^a n_3^b F_{ST,23} - \d^{ab} F_{ST, \d},
\end{align}
with
\begin{align}
F_{ST, 32} = F_{ST, 23}, \qquad F_{ST, \d} = \f{1}{d} \l ( F_{ST,33} + F_{ST,22} + 2 F_{ST,32} \cos \t \r),
\end{align}

\paragraph{Vector-tensor}

\begin{align}
G^{abc}_{JT} & = \sum_{i,j,k=2}^3 n_i^a n_j^b n_k^c \, F_{JT, ijk} \nn
+ \sum_{i=2}^3 \l ( n_i^b \d^{ac} + n_i^c \d^{ab} \r ) F_{JT, \d i} - \d^{bc} \sum_{i=2}^3 n_i^a F_{JT, i \d},
\end{align}
with
\begin{align}
F_{JT, 3\d} & =\f{1}{d} \l (F_{JT, 333}+F_{JT, 322} + 2 F_{JT, \d3} + 2 F_{JT, 332} \cos \t \r ), \\
F_{JT, 2\d} & =\f{1}{d} \l (F_{JT, 222}+F_{JT, 233} + 2 F_{JT, \d2} + 2 F_{JT, 232} \cos \t \r ), \\
F_{JT, ijk} & = F_{JT, ikj} 
\end{align}

\paragraph{Tensor-tensor}

\begin{align}
G^{abcd}_{TT} & = 
\sum_{i,j,k,\ell=2}^3 n_i^a n_j^b n_k^c n_\ell^d \, F_{TT, ijk\ell} \\
& \quad +
\f{1}{4} \sum_{i,j=2}^3 \l ( \d^{ac} n_i^b n_j^d + \d^{ad} n_i^b n_j^c + \d^{bc} n_i^a n_j^d + \d^{bd} n_i^a n_j^c \r ) F_{TT, i \d j} \nn \\
& \quad +
\f{1}{2} \l ( \d^{ac} \d^{bd} + \d^{ad} \d^{bc} \r )F_{TT, \d \d} \nn \\
& \qquad -
\d^{ab} \sum_{i,j=2}^3 n_i^c n_j^d F_{TT, \d i j} -\d^{cd} \sum_{i,j=2}^3 n_i^a n_j^b F_{TT, i j \d} + F_{TT,0} \d^{ad} \d^{bc} \nn \\
F_{TT, ijk\ell} & = F_{TT, j i k\ell} = F_{TT, ij\ell k}, \quad F_{TT, \d ij} = F_{TT, \d ji}, \quad F_{TT, i j \d} = F_{TT, j i \d},
\end{align}
with
\begin{align}
F_{TT,\delta 22}
&=
\frac{1}{d}
\left(
F_{TT,2\delta 2}
+\cos\theta\, F_{TT,2322}
+\cos\theta\, F_{TT,3222}
+F_{TT,2222}
+F_{TT,3322}
\right),
\\
F_{TT,22\delta}
&=
\frac{1}{d}
\left(
F_{TT,2\delta 2}
+\cos\theta\, F_{TT,2223}
+\cos\theta\, F_{TT,2232}
+F_{TT,2222}
+F_{TT,2233}
\right),
\end{align}

\begin{align}
F_{TT,\delta 32}
&=
\frac{1}{d}
\left[
\frac{1}{2}
\left(
F_{TT,2\delta 3}
+F_{TT,3\delta 2}
\right)
+\cos\theta
\left(
F_{TT,2332}
+F_{TT,3232}
\right)
+F_{TT,2232}
+F_{TT,3332}
\right],
\\
F_{TT,32\delta}
&=
\frac{1}{d}
\left[
\frac{1}{2}
\left(
F_{TT,2\delta 3}
+F_{TT,3\delta 2}
\right)
+\cos\theta
\left(
F_{TT,3223}
+F_{TT,3232}
\right)
+F_{TT,3222}
+F_{TT,3233}
\right],
\end{align}

\begin{align}
F_{TT,\delta 33}
&=
\frac{1}{d}
\left(
F_{TT,3\delta 3}
+\cos\theta\, F_{TT,2333}
+\cos\theta\, F_{TT,3233}
+F_{TT,2233}
+F_{TT,3333}
\right),
\\
F_{TT,33\delta}
&=
\frac{1}{d}
\left(
F_{TT,3\delta 3}
+\cos\theta\, F_{TT,3323}
+\cos\theta\, F_{TT,3332}
+F_{TT,3322}
+F_{TT,3333}
\right),
\end{align}

\begin{align}
F_{TT,0}
&=
\frac{1}{d^2}
\Bigg[
\left(
-d F_{TT,\delta\delta}
+F_{TT,2\delta 2}
+F_{TT,3\delta 3}
+F_{TT,2222}
+F_{TT,2233}
+F_{TT,3322}
+F_{TT,3333}
\right)
\nonumber\\
&\hspace{2.5cm}
+
\Big (
F_{TT,2\delta 3}
+F_{TT,3\delta 2}
+2F_{TT,2232}
+2F_{TT,3222}
+2F_{TT,3233}
+2F_{TT,3332}
 \Big ) \cos\theta
\nonumber\\
&\hspace{3cm}
+4F_{TT,3232} \cos^2\theta
\Bigg].
\end{align}

\subsection{Primary and first-descendant contributions in the \(F\)-basis}
\label{app:F-descendants}

In this subsection we list the contribution of the first descendant of the leading exchanged primary to the \(F\)-basis coefficient functions. Trace contributions are tacitly assumed but are not presented explicitly. To keep the presentation compact, we list only the non-vanishing coefficient functions. We factor out the leading OPE coefficients, Ward-identity factors, and the leading charge-dependent powers of \(|z|\). Thus the expressions below give only the relative descendant contribution. We also use
\begin{equation}
\Delta_Q=\alpha |Q|^{3/2}+\cdots .
\end{equation}
Whenever the descendant normalization depends only on the magnitude of the large charge, we write the suppression in terms of \(|Q|\). This is the case for the \(SS\), \(JJ\), \(JT\), and \(TT\) channels. In the mixed charged-scalar channels \(SJ\) and \(ST\), however, the descendant coefficient is linear in the difference between the dimensions of the two neighboring charge sectors,
\begin{equation}
\Delta_{Q+\frac q2}-\Delta_{Q-\frac q2}
=
q\,\partial_Q\Delta_Q+\cdots .
\end{equation}
Since \(\partial_Q\Delta_Q\) changes sign under \(Q\to -Q\), the natural suppression factor in these channels is proportional to \(1/Q\), not \(1/|Q|\). This sign is important when applying crossing.\\
The quantities displayed below are the relative functions \(F^{(Q),1}_{AB,r}\), not the full coefficient functions. Overall leading OPE coefficients, Ward-identity factors, and leading powers of \(|z|\) have therefore already been divided out.

\paragraph{Scalar-scalar}

\begin{equation}
F_{SS}^{(Q),1}
=
\alpha\,
\frac{9q^2}{8\sqrt{|Q|}}\,
e^\tau\,
\cos\theta .
\end{equation}

\paragraph{Scalar-current}

\begin{equation}
\begin{alignedat}{2}
F_{SJ,2}^{(Q),1}
&=
\frac{3q}{4Q}\,
e^\tau\,
\cos\theta,
&\qquad
F_{SJ,3}^{(Q),1}
&=
\frac{3q}{4Q}\,
e^\tau .
\end{alignedat}
\end{equation}

\paragraph{Current-current}

\begin{equation}
\begin{alignedat}{2}
F_{JJ,33}^{(Q),1}
&=
-\frac{3}{2\alpha |Q|^{3/2}}\,
e^\tau,
&\qquad
F_{JJ,32}^{(Q),1}
&=
-\frac{3}{2\alpha |Q|^{3/2}}\,
e^\tau\,
\cos\theta,
\\
F_{JJ,22}^{(Q),1}
&=
\frac{1}{2\alpha |Q|^{3/2}}\,
e^\tau,
&\qquad
F_{JJ,\delta}^{(Q),1}
&=
\frac{1}{2\alpha |Q|^{3/2}}\,
e^\tau .
\end{alignedat}
\end{equation}

\paragraph{Scalar-tensor}

\begin{equation}
\begin{alignedat}{2}
F_{ST,23}^{(Q),1}
&=
\frac{3q}{4Q}\,
e^\tau,
&\qquad
F_{ST,22}^{(Q),1}
&=
\frac{3q}{4Q}\,
e^\tau\,
\cos\theta .
\end{alignedat}
\end{equation}

\paragraph{Current-tensor}

\begin{equation}
\begin{alignedat}{2}
F_{JT,332}^{(Q),1}
&=
-\frac{3}{2\alpha |Q|^{3/2}}\,
e^\tau,
&\qquad
F_{JT,322}^{(Q),1}
&=
-\frac{3}{2\alpha |Q|^{3/2}}\,
e^\tau\,
\cos\theta,
\\
F_{JT,222}^{(Q),1}
&=
\frac{1}{2\alpha |Q|^{3/2}}\,
e^\tau,
&\qquad
F_{JT,\delta2}^{(Q),1}
&=
\frac{1}{2\alpha |Q|^{3/2}}\,
e^\tau .
\end{alignedat}
\end{equation}

\paragraph{Tensor-tensor}

\begin{equation}
\begin{alignedat}{2}
F_{TT,3332}^{(Q),1}
&=
\frac{5}{2\alpha |Q|^{3/2}}\,
e^\tau,
&\qquad
F_{TT,3322}^{(Q),1}
&=
\frac{5}{2\alpha |Q|^{3/2}}\,
e^\tau\,
\cos\theta,
\\
F_{TT,3222}^{(Q),1}
&=
-\frac{1}{2\alpha |Q|^{3/2}}\,
e^\tau,
&\qquad
F_{TT,3\delta2}^{(Q),1}
&=
-\frac{2}{\alpha |Q|^{3/2}}\,
e^\tau .
\end{alignedat}
\end{equation}

\section{Relation between the \texorpdfstring{\(F\)}{F}- and \texorpdfstring{\(H\)}{H}-bases \label{app:F-to-H}}

In this appendix we relate the \(F\)-basis used in Appendix~\ref{app:F-basis} to the \(H\)-basis used in the main text. The \(H\)-basis is obtained by taking linear combinations of the \(F\)-basis structures with coefficients depending on \(\cos\t\). As in Appendix~\ref{app:F-basis}, all coefficient functions depend on \((\tau,\cos\t)\), but we suppress these arguments in order to keep the formulas readable. The advantage of the \(H\)-basis is that the resulting functions have simple crossing properties and admit a natural organization by levels. This organization is especially useful when deriving the algebraic bootstrap equations.

\paragraph{Scalar-scalar}

\begin{equation}
H_{SS} =F_{SS}
\end{equation}

\paragraph{Scalar-current}

\begin{align}
H_{SJ,3} & = F_{SJ,3}, \\
H_{SJ,0} & = F_{SJ,2} + F_{SJ,3} \cos \t, 
\end{align}

\paragraph{Current-current}

\begin{align}
H_{JJ,23} & = F_{JJ,23}, \\
H_{JJ,\d} & = F_{JJ,\d}, \\
-H_{JJ,3} & = F_{JJ,33} + F_{JJ,\d} + F_{JJ,23} \cos \t, \\
H_{JJ,2} & = F_{JJ,22} + F_{JJ,\d} + F_{JJ,23} \cos \t, \\
-H_{JJ,0} & = F_{JJ,32}+ \l ( F_{JJ,22} + F_{JJ,\d} + F_{JJ,33} \r ) \cos \t + F_{JJ,23} \cos^2 \t ,
\end{align}

\paragraph{Scalar-tensor}

\begin{align}
-H_{ST,33} & = F_{ST,33}, \\
H_{ST,3} & = F_{ST,32} + F_{ST,33} \cos \t, \\
H_{ST,0} & = \f{d-1}{d} F_{ST,22} - \f{1}{d} F_{ST,33} + 2 \f{d-1}{d} F_{ST,32} \cos \t
+ F_{ST,33} \cos^2 \t ,
\end{align}

\paragraph{Current-tensor}

\begin{align}
\text{Level 0}: \quad 
-H_{JT, 233} & =F_{JT, 233} \\
-H_{JT, \d3} & = F_{JT, \d3} \\
\text{Level 1}: \quad 
H_{JT, 23} & = F_{JT, 232} + F_{JT, \d 3}+ F_{JT, 233} \cos \t \\
H_{JT, \d} & = F_{JT, \d2} + F_{JT, \d3} \cos \t \\
H_{JT, 33} & = F_{JT, 333} + 2F_{JT, \d 3}+ F_{JT, 233} \cos \t
\end{align}

\begin{align}
\text{Level 2}: \quad
-H_{JT, 3} & = F_{JT, 332} + F_{JT, \d 2}+ \l ( F_{JT, 333} + F_{JT, 232} + 2 F_{JT, \d3}\r )\cos \t + F_{JT, 233} \cos^2 \t \\
H_{JT, 2} & =  
\f{d-1}{d} \l (F_{JT, 222} + 2 F_{JT, \d 2}\r ) - \f{1}{d} F_{JT, 233} \nn \\
& \quad
+ \l [ 2 F_{JT, \d3} + \f{d-1}{d} \l (F_{JT, 223} + F_{JT, 232} \r ) \r ]\cos \t + F_{JT, 233} \cos^2 \t
\end{align}

\begin{align}
\text{Level 3}: \quad
-H_{JT, 0} & =
\frac{1}{d} \l [ (d-1) F_{322}-2 F_{\delta 3}-F_{333} \r ] \\
& \quad + 
\frac{1}{d}\l [2 (d-1) F_{\delta 2}+(d-1) F_{222}+(d-1) \left(F_{323}+F_{332} \right )-F_{233}\r ]\cos \t \nn \\
& \quad + 
\l [\frac{d-1}{d} \left(F_{223}+F_{232}\right)+2 F_{\delta 3}+F_{333} \r ] \cos^2 \t + F_{233} \cos^3 \t \nn
\end{align}

\paragraph{Tensor-tensor}

\begin{align}
\text{Level 0}: \quad 
H_{TT, 2233} & =F_{TT, 2233} \\
H_{TT, 2\d3} & =F_{TT, 2\d3} \\
H_{TT, \d\d} & =F_{TT, \d\d}
\end{align}

\begin{align}
\text{Level 1}: \quad
-H_{TT, 223} & = \f{1}{2} F_{TT, 2\d3} + F_{TT, 2232} + F_{TT, 2233}\cos\t \\
-H_{TT, 2\d} & = \f{1}{2} F_{TT, \d\d} + \f{1}{4}F_{TT, 2\d2} + \f{1}{4}F_{TT, 2\d3}\cos\t \\
H_{TT, 233} & = \f{1}{2} F_{TT, 2\d3} + F_{TT, 2233} + F_{TT, 3233}\cos\t \\
H_{TT, \d3} & = \f{1}{2} F_{TT, \d\d} + \f{1}{4}F_{TT, 3\d3} + \f{1}{4}F_{TT, 2\d3}\cos\t
\end{align}

\begin{align}
\text{Level 2}: \quad
-H_{TT,33}
&=
F_{TT,\delta\delta}
+\left(1-\frac{1}{d}\right)F_{TT,3\delta 3}
-\frac{1}{d}F_{TT,2233}
+\left(1-\frac{1}{d}\right)F_{TT,3333}
\nonumber\\
&\quad
+\cos\theta
\left[
F_{TT,2\delta 3}
+2\left(1-\frac{1}{d}\right)F_{TT,3233}
\right]
+\cos^2\theta\, F_{TT,2233} \\
-H_{TT,22}
&=
F_{TT,\delta\delta}
+\left(1-\frac{1}{d}\right)F_{TT,2\delta 2}
+\left(1-\frac{1}{d}\right)F_{TT,2222}
-\frac{1}{d}F_{TT,2233}
\nonumber\\
&\quad
+\cos\theta
\left[
F_{TT,2\delta 3}
+2\left(1-\frac{1}{d}\right)F_{TT,2232}
\right]
+\cos^2\theta\,F_{TT,2233},
\\
-H_{TT,23}
&=
\frac{1}{2}F_{TT,\delta\delta}
+\frac{1}{4}F_{TT,2\delta 2}
+\frac{1}{4}F_{TT,3\delta 3}
+F_{TT,3232}
\nonumber\\
&\quad
+\cos\theta
\left(
\frac{3}{4}F_{TT,2\delta 3}
+F_{TT,2232}
+F_{TT,3233}
\right)
+\cos^2\theta\,F_{TT,2233},
\\
-H_{TT,\delta}
&=
\frac{1}{4}F_{TT,3\delta 2}
+\cos\theta
\left(
\frac{1}{2}F_{TT,\delta\delta}
+\frac{1}{4}F_{TT,2\delta 2}
+\frac{1}{4}F_{TT,3\delta 3}
\right)
+\frac{1}{4}\cos^2\theta\,F_{TT,2\delta 3}
\end{align}

\begin{align}
\text{Level 3}: \quad 
H_{TT,3}
&=
-\frac{1}{2d}F_{TT,2\delta 3}
+\frac{1}{2}\left(1-\frac{1}{d}\right)F_{TT,3\delta 2}
-\frac{1}{d}F_{TT,2232}
+\left(1-\frac{1}{d}\right)F_{TT,3332}
\nonumber\\
&\quad
+
\bigg [
F_{TT,\delta\delta}
+\frac{1}{2}F_{TT,2\delta 2}
-\frac{1}{d}F_{TT,2233}
+\left(1-\frac{1}{d}\right) \Big (F_{TT,3\delta 3}
+2F_{TT,3232}+F_{TT,3333} \Big )
\bigg ]
\cos\theta
\nonumber\\
&\quad
+
\left[
F_{TT,2\delta 3}
+F_{TT,2232}
+2\left(1-\frac{1}{d}\right)F_{TT,3233}
\right]
\cos^2\theta
+F_{TT,2233} \cos^3\theta,
\\
-H_{TT,2}
&=
-\frac{1}{2d}F_{TT,2\delta 3}
+\frac{1}{2}\left(1-\frac{1}{d}\right)F_{TT,3\delta 2}
+\left(1-\frac{1}{d}\right)F_{TT,3222}
-\frac{1}{d}F_{TT,3233}
\nonumber\\
&\quad
+
\bigg [
F_{TT,\delta\delta}
+\frac{1}{2}F_{TT,3\delta 3}
-\frac{1}{d}F_{TT,2233}
+\left(1-\frac{1}{d}\right)
\Big ( 
F_{TT,2\delta 2} + F_{TT,2222} + 2 F_{TT,3232} \Big )
\bigg ]
\cos\theta
\nonumber\\
&\quad
+
\left[
F_{TT,2\delta 3}
+2\left(1-\frac{1}{d}\right)F_{TT,2232}
+F_{TT,3233}
\right]
\cos^2\theta
+F_{TT,2233} \cos^3\theta.
\end{align}

\begin{align}
\text{Level 4}: \quad 
H_{TT,0}
&=
-\frac{1}{d}F_{TT,\delta\delta}
-\frac{d-1}{d^2}F_{TT,2\delta 2}
-\frac{d-1}{d^2}F_{TT,3\delta 3}
-\frac{d-1}{d^2}F_{TT,2222}
+\frac{1}{d^2}F_{TT,2233}
\nonumber\\
&\quad
+\left(1-\frac{1}{d}\right)^2F_{TT,3322}
-\frac{d-1}{d^2}F_{TT,3333}
\nonumber\\
&\quad
+
\left[
-\frac{2d-1}{d^2}F_{TT,2\delta 3}
+\left(1-\frac{1}{d}\right)^2 F_{TT,3\delta 2}
-2\frac{d-1}{d^2}F_{TT,2232}
\right.
\nonumber\\
&\hspace{3.0cm}
\left.
+2\left(1-\frac{1}{d}\right)^2F_{TT,3222}
-\frac{2d-2}{d^2}F_{TT,3233}
+2\left(1-\frac{1}{d}\right)^2F_{TT,3332}
\right]
\cos\theta
\nonumber\\
&\quad
+
\Bigg [
F_{TT,\delta\delta}
-\frac{2}{d}F_{TT,2233}
+\left(1-\frac{1}{d}\right)F_{TT,2\delta 2}
+\left(1-\frac{1}{d}\right)F_{TT,3\delta 3}
\nonumber\\
&\hspace{3.0cm}
+4\left(1-\frac{1}{d}\right)^2F_{TT,3232}
+\left(1-\frac{1}{d}\right)F_{TT,3333}
+\left(1-\frac{1}{d}\right)F_{TT,2222}
\Bigg ] 
\cos^2\theta
\nonumber\\
&\quad
+
\left[
F_{TT,2\delta 3}
+2\left(1-\frac{1}{d}\right)F_{TT,2232}
+2\left(1-\frac{1}{d}\right)F_{TT,3233}
\right]
\cos^3\theta
+F_{TT,2233} \cos^4\theta.
\end{align}

\subsection{First descendant contribution in the \texorpdfstring{\(H\)}{H}-basis \label{app:H-descendants}}

Applying the \(F\)-to-\(H\) map to the descendant contribution in Appendix~\ref{app:F-descendants}, one obtains the following non-vanishing \(H^{(1)}\)-functions. As in Appendix~\ref{app:F-descendants}, suppressions that depend only on the magnitude of the large charge are written in terms of \(|Q|\). In the mixed charged-scalar channels \(SJ\) and \(ST\), the suppression is kept as \(1/Q\) since it changes sign under \(Q\to -Q\).

\paragraph{Scalar-scalar}

\begin{equation}
H_{SS}^{(Q),1}
=
\alpha\,
\frac{9q^2}{8\sqrt{|Q|}}\,
e^\tau\,
\cos\theta .
\end{equation}

\paragraph{Scalar-current}

\begin{equation}
\begin{alignedat}{2}
H_{SJ,3}^{(Q),1}
&=
\frac{3q}{4Q}\,
e^\tau,
&\qquad
H_{SJ,0}^{(Q),1}
&=
\frac{3q}{2Q}\,
e^\tau\,
\cos\theta .
\end{alignedat}
\end{equation}

\paragraph{Current-current}

\begin{equation}
\begin{alignedat}{2}
H_{JJ,\delta}^{(Q),1}
&=
\frac{1}{2\alpha |Q|^{3/2}}\,
e^\tau,
&\qquad
H_{JJ,3}^{(Q),1}
&=
\frac{1}{\alpha |Q|^{3/2}}\,
e^\tau,
\\
H_{JJ,2}^{(Q),1}
&=
\frac{1}{\alpha |Q|^{3/2}}\,
e^\tau,
&\qquad
H_{JJ,0}^{(Q),1}
&=
\frac{2}{\alpha |Q|^{3/2}}\,
e^\tau\,
\cos\theta .
\end{alignedat}
\end{equation}

\paragraph{Scalar-tensor}

\begin{equation}
\begin{alignedat}{2}
H_{ST,3}^{(Q),1}
&=
\frac{9q}{8Q}\,
e^\tau,
&\qquad
H_{ST,0}^{(Q),1}
&=
\frac{9q}{4Q}\,
e^\tau\,
\cos\theta .
\end{alignedat}
\end{equation}

\paragraph{Current-tensor}

\begin{equation}
\begin{alignedat}{2}
H_{JT,\delta}^{(Q),1}
&=
\frac{3}{4\alpha |Q|^{3/2}}\,
e^\tau,
&\qquad
H_{JT,3}^{(Q),1}
&=
\frac{3}{2\alpha |Q|^{3/2}}\,
e^\tau,
\\
H_{JT,2}^{(Q),1}
&=
\frac{3}{2\alpha |Q|^{3/2}}\,
e^\tau,
&\qquad
H_{JT,0}^{(Q),1}
&=
\frac{3}{\alpha |Q|^{3/2}}\,
e^\tau\,
\cos\theta .
\end{alignedat}
\end{equation}

\paragraph{Tensor-tensor}

\begin{equation}
\begin{alignedat}{2}
H_{TT,\delta}^{(Q),1}
&=
\frac{9}{8\alpha |Q|^{3/2}}\,
e^\tau,
&\qquad
H_{TT,3}^{(Q),1}
&=
\frac{9}{4\alpha |Q|^{3/2}}\,
e^\tau,
\\
H_{TT,2}^{(Q),1}
&=
\frac{9}{4\alpha |Q|^{3/2}}\,
e^\tau,
&\qquad
H_{TT,0}^{(Q),1}
&=
\frac{9}{2\alpha |Q|^{3/2}}\,
e^\tau\,
\cos\theta .
\end{alignedat}
\end{equation}
One small point requires care in channels involving the stress tensor. The corresponding \(H\)-basis functions are defined with additional numerical factors. These factors are part of the change of basis from \(F\) to \(H\). They are therefore present both in the leading primary contribution and in the first-descendant contribution. Since \(H^{(Q),1}_{AB,r}\) is defined as a relative correction after dividing by \(H^{(Q),0}_{AB,0}\), these basis-conversion factors cancel and should not be included again in the relative descendant coefficient.

\section{Three-point structures and conservation constraints\label{app:three-point-structures}}

In this appendix we summarize the three-point structures that determine which primary states can appear in the spinning channels. We use the parametrization of Refs.~\cite{Costa:2011mg,Costa:2011dw}. A conformal partial wave is generated by differential operators acting on a seed structure,
\begin{align}
\label{eq:DifferentialBasis}
\left\{
\begin{array}{ccc}
\D_1 & \D_2 & \D_0 \\
\ell_1 & \ell_2 & \ell_0 \\
n_{20} & n_{10} & n_{12}
\end{array}
\right\}
&=
H_{12}^{n_{12}}
D_{12}^{n_{10}}
D_{21}^{n_{20}}
D_{11}^{m_1}
D_{22}^{m_2}
\left[
\begin{array}{ccc}
\D_1+m_1+n_{20}+n_{12}
&
\D_2+m_2+n_{10}+n_{12}
&
\D_0
\\
0 & 0 & \ell_0 \\
0 & 0 & 0
\end{array}
\right].
\end{align}
The corresponding four-point function is written as
\begin{equation}
{\cal D}_{\rm left}\,
{\cal D}_{\rm right}\,
W_{\cal O}(P_4,P_3,P_2,P_1),
\end{equation}
with
\begin{equation}
{\cal D}_{\rm right}
=
H_{12}^{n_{12}}
D_{12}^{n_{10}}
D_{21}^{n_{20}}
D_{11}^{m_1}
D_{22}^{m_2}
\Sigma^{m_1+n_{20}+n_{12},\,m_2+n_{10}+n_{12}},
\end{equation}
and
\begin{equation}
{\cal D}_{\rm left}
=
{\cal D}_{\rm right}(1\leftrightarrow 4,\,2\leftrightarrow 3).
\end{equation}
Let us also recall the counting of independent three-point structures in our case. We consider
\begin{equation}
\left\langle
{\cal O}_{a_1\dots a_\ell}(x_3)\,
{\cal A}(x_2)\,
\Phi(x_1)
\right\rangle ,
\end{equation}
where
\begin{equation}
{\cal A}(x_2)=J_a(x_2),\qquad T_{ab}(x_2).
\end{equation}
The possible structures are
\begin{align}
J:
&\qquad
V_3^\ell V_2,
\qquad
V_3^{\ell-1}H_{23},
\\
T:
&\qquad
V_3^\ell V_2,
\qquad
V_3^{\ell-1}V_2H_{23},
\qquad
V_3^{\ell-2}H_{23}^2 .
\end{align}
Therefore the number of structures is
\begin{equation}
\begin{array}{c|ccc}
& \ell\geq 2 & \ell=1 & \ell=0 \\
\hline
J & 2 & 2 & 1 \\
T & 3 & 2 & 1
\end{array}
\end{equation}

In our case, the couplings are labelled by the number of times the operator at point \(2\) is contracted with the exchanged operator. For the stress tensor, the following parametrization applies for \(\ell\geq2\), while for the current it applies for \(\ell\geq1\):
\begin{align}
&{\cal O}_{\ell} S \Phi:
\qquad
\lambda_S,
&&
\Phi^\dagger S {\cal O}_{\ell}:
\qquad
\lambda_S^*,
\\
&{\cal O}_{\ell} J \Phi:
\qquad
\lambda_{1,0},\lambda_{1,1},
&&
\Phi^\dagger J {\cal O}_{\ell}:
\qquad
\lambda_{1,0}^*,\lambda_{1,1}^*,
\\
&{\cal O}_{\ell} T \Phi:
\qquad
\lambda_{2,0},\lambda_{2,1},\lambda_{2,2},
&&
\Phi^\dagger T {\cal O}_{\ell}:
\qquad
\lambda_{2,0}^*,\lambda_{2,1}^*,\lambda_{2,2}^* .
\end{align}
Current conservation gives, for \(\ell\geq 1\),
\begin{equation}
\lambda_{1,0}
=
\frac{
\left(\Delta_1-\Delta_0+\ell\right)
\left(d-\Delta_1+\Delta_0+\ell-2\right)
}{
d\Delta_1-d\Delta_0+\ell(d+\ell-2)-\Delta_1^2+\Delta_0^2
}
\lambda_{1,1}.
\end{equation}
Energy-momentum tensor conservation gives, for \(\ell\geq 2\),
\begin{align}
\lambda_{2,1}
&=
\lambda_{2,0}
\frac{
2\left(
-d\Delta_0
+
\left(\Delta_1+\ell-2\right)
\left(d-\Delta_1+\ell\right)
+
\Delta_0^2
\right)
}{
\left(\Delta_1-\Delta_0+\ell-2\right)
\left(d-\Delta_1+\Delta_0+\ell\right)
},
\\
\lambda_{2,2}
&=
\frac{\lambda_{2,0}}{
(d-1)
\left(\Delta_1-\Delta_0+\ell-2\right)
\left(\Delta_1-\Delta_0+\ell\right)
\left(d-\Delta_1+\Delta_0+\ell-2\right)
\left(d-\Delta_1+\Delta_0+\ell\right)
}
\nonumber\\
&\quad \times
\bigg\{
(d-1)\Delta_1^2
\Big[
d^2
+
2\Delta_0(d-\Delta_0)
-
2d(\ell-1)
-
2(\ell-2)\ell
-
4
\Big]
\nonumber\\
&\qquad
+
\Delta_0(d-\Delta_0)
\Big[
-2d^2(\ell-1)
+
(d-1)\Delta_0(d-\Delta_0)
-
2d\big((\ell-5)\ell+3\big)
+
6(\ell-2)\ell
+
4
\Big]
\nonumber\\
&\qquad
+
(d-1)\Delta_1^4
-
2(d-1)d\Delta_1^3
\nonumber\\
&\qquad
+
2(d-1)d\Delta_1
\Big[
\Delta_0(\Delta_0-d)
+
d(\ell-1)
+
(\ell-2)\ell
+
2
\Big]
\nonumber\\
&\qquad
+
(d-1)(\ell-2)\ell(d+\ell-2)(d+\ell)
\bigg\}.
\end{align}
In writing these relations, we tacitly assume that the denominators do not vanish. Exceptional cases in which a denominator vanishes must be treated separately.

For \(\ell=1\) in the tensor case, there are only two independent structures. A nonzero solution of the conservation equations only exists for\footnote{More generally, one finds \(\Delta_0=\Delta_1\pm 1\), but we assume \(\Delta_0\geq \Delta_1\).}
\begin{equation}
\label{eq:l1Exchange}
\Delta_0=\Delta_1+1.
\end{equation}
In this case,
\begin{equation}
\lambda_{2,1}
=
\lambda_{2,0}
\frac{2(d+2-2\Delta_0)}{d+2}.
\end{equation}
For these parameters one also has
\begin{equation}
\lambda_{1,0}=0.
\end{equation}
For \(\ell=0\), correlators involving either the current or the stress tensor are nonzero only if
\begin{equation}
\label{eq:l0Exchange}
\Delta_0=\Delta_1.
\end{equation}

\section{Spectral sums for $\ell \geq 2$}
\label{app:spectral-decomposition}

To derive the spectral sums over Gegenbauer polynomials we insert a complete set of charge-\(Q\) primary states between the two probe operators on the cylinder. A primary of spin \(\ell\) gives a traceless symmetric tensor representation of \(SO(3)\), ${\mc O}_{a_1\dots a_\ell}$. Its contribution to a cylinder two-point function is proportional to
\begin{equation}
e^{\omega_\ell\tau},
\end{equation}
where
\begin{equation}
\omega_\ell=\Delta_{\cal O}-\Delta_Q
\end{equation}
is the excitation energy above the charge-\(Q\) ground state.

For scalar probes, the angular dependence is fixed by rotational invariance. The sum over the polarizations of a spin-\(\ell\) traceless symmetric tensor gives the standard addition theorem on the sphere. In \(d=3\), this gives
\begin{equation}
\sum_{\text{pol.}}
{\cal O}_{a_1\dots a_\ell}(n_3)
{\cal O}_{a_1\dots a_\ell}(n_2)
\propto
C_\ell^{(1/2)}(\cos\theta).
\end{equation}
Thus the scalar-scalar contribution takes the form
\begin{equation}
\sum_{\ell}
|\lambda_S|^2
e^{\omega_\ell\tau}
C_\ell^{(1/2)}(\cos\theta).
\end{equation}

For spinning probes, the tensor structures in the three-point functions contain contractions between the external spin indices and the exchanged spin-\(\ell\) operator. After projecting onto the \(H\)-basis, these contractions act on the scalar Gegenbauer structure by effectively lowering the spin label and increasing the Gegenbauer index. This follows from the identity
\begin{equation}
\frac{d}{dx}C_\ell^{(\alpha)}(x)
=
2\alpha\, C_{\ell-1}^{(\alpha+1)}(x).
\end{equation}
Repeated contractions therefore lead to the pattern
\begin{equation}
C_\ell^{(1/2)}(\cos\theta)
\longrightarrow
C_{\ell-1}^{(3/2)}(\cos\theta),
\quad
C_{\ell-2}^{(5/2)}(\cos\theta),
\quad
C_{\ell-3}^{(7/2)}(\cos\theta),
\quad \ldots .
\end{equation}
Consequently, each \(H\)-basis coefficient function receives primary-state contributions of the schematic form
\begin{equation}
h_{AB,\alpha}^{\rm prim}
=
\sum_{\ell}
\lambda_A^*\lambda_B\,
{\cal K}_{AB,\alpha}(\omega_\ell,J_\ell^2)\,
e^{\omega_\ell\tau}
C_{\ell-k_\alpha}^{(1/2+k_\alpha)}(\cos\theta),
\end{equation}
where \(k_\alpha\) (consequently, the lower limits as well) is determined by the tensor structure. The functions \({\cal K}_{AB,\alpha}\) are rational functions of \(\omega_\ell\) and \(J_\ell^2\). They are fixed by the three-point tensor structures, conservation of \(J\) and \(T\), and the projection from the \(F\)-basis to the \(H\)-basis.

In practice, computing these contributions by multiplying explicit tensor structures is rather inefficient, especially in channels involving more than one spinning probe. Instead, we use the method of Refs.~\cite{Costa:2011mg,Costa:2011dw}. In this approach, spinning conformal partial waves are generated by acting with differential operators in embedding space on scalar conformal partial waves. We recalled this differential basis in Appendix~\ref{app:three-point-structures}, see in particular Eq.~\eqref{eq:DifferentialBasis}. After acting with the appropriate differential operators, we project to physical space and identify the radial dependence
\begin{equation}
|z|^{\omega_\ell},
\end{equation}
and the tensor structure. This gives the corresponding contributions to the \(F\)-basis functions, and the \(H\)-basis functions are then obtained by the change of basis described in Appendix~\ref{app:F-to-H}.

Let us also recall the counting of independent structures. As explained in Appendix~\ref{app:three-point-structures}, a three-point function involving a scalar probe has one structure, a three-point function involving the current has two structures, and a three-point function involving the stress tensor has three structures. Therefore the number of independent structures appearing in the six four-point sectors is
\begin{equation}
\begin{array}{c|cccccc}
\text{sector}
&
SS
&
SJ
&
JJ
&
ST
&
JT
&
TT
\\
\hline
\text{number of structures}
&
1
&
1\times 2=2
&
2\times 2=4
&
1\times 3=3
&
2\times 3=6
&
3\times 3=9
\end{array}
\end{equation}

\section{Crossing equations \label{app:CrossingHBasis}}

In this appendix we collect the crossing parities of the individual coefficient functions. The origin of the crossing equations and the derivation of the corresponding algebraic bootstrap constraints are discussed in Section~\ref{sec:bootstrap-equations}. Here we only record the transformation properties of $H$- and $h$-functions.

\subsection{Crossing equations for the \(H\)-basis \label{app:CrossingH-functions}}

\paragraph{Scalar-scalar}

\begin{equation}
H^{(-Q)}_{SS}(-\tau,\cos\theta)
=
H^{(Q)}_{SS}(\tau,\cos\theta).
\end{equation}

\paragraph{Scalar-current}

\begin{equation}
H^{(-Q)}_{SJ,3}(-\tau,\cos\theta)
=
H^{(Q)}_{SJ,3}(\tau,\cos\theta),
\end{equation}
and
\begin{equation}
H^{(-Q)}_{SJ,0}(-\tau,\cos\theta)
=
-
H^{(Q)}_{SJ,0}(\tau,\cos\theta).
\end{equation}

\paragraph{Current-current}

\begin{align}
H^{(-Q)}_{JJ,23}(-\tau,\cos\theta)
&=
H^{(Q)}_{JJ,23}(\tau,\cos\theta),
\\
H^{(-Q)}_{JJ,\delta}(-\tau,\cos\theta)
&=
H^{(Q)}_{JJ,\delta}(\tau,\cos\theta),
\\
H^{(-Q)}_{JJ,0}(-\tau,\cos\theta)
&=
H^{(Q)}_{JJ,0}(\tau,\cos\theta),
\end{align}
and
\begin{align}
H^{(-Q)}_{JJ,3}(-\tau,\cos\theta)
&=
-
H^{(Q)}_{JJ,3}(\tau,\cos\theta),
\\
H^{(-Q)}_{JJ,2}(-\tau,\cos\theta)
&=
-
H^{(Q)}_{JJ,2}(\tau,\cos\theta).
\end{align}

\paragraph{Scalar-tensor}

\begin{align}
H^{(-Q)}_{ST,33}(-\tau,\cos\theta)
&=
H^{(Q)}_{ST,33}(\tau,\cos\theta),
\\
H^{(-Q)}_{ST,0}(-\tau,\cos\theta)
&=
H^{(Q)}_{ST,0}(\tau,\cos\theta),
\end{align}
and
\begin{equation}
H^{(-Q)}_{ST,3}(-\tau,\cos\theta)
=
-
H^{(Q)}_{ST,3}(\tau,\cos\theta).
\end{equation}

\paragraph{Current-tensor}

\begin{align}
H^{(-Q)}_{JT,233}(-\tau,\cos\theta)
&=
H^{(Q)}_{JT,233}(\tau,\cos\theta),
\\
H^{(-Q)}_{JT,\delta3}(-\tau,\cos\theta)
&=
H^{(Q)}_{JT,\delta3}(\tau,\cos\theta),
\\
H^{(-Q)}_{JT,2}(-\tau,\cos\theta)
&=
H^{(Q)}_{JT,2}(\tau,\cos\theta),
\\
H^{(-Q)}_{JT,3}(-\tau,\cos\theta)
&=
H^{(Q)}_{JT,3}(\tau,\cos\theta),
\end{align}
and
\begin{align}
H^{(-Q)}_{JT,23}(-\tau,\cos\theta)
&=
-
H^{(Q)}_{JT,23}(\tau,\cos\theta),
\\
H^{(-Q)}_{JT,\delta}(-\tau,\cos\theta)
&=
-
H^{(Q)}_{JT,\delta}(\tau,\cos\theta),
\\
H^{(-Q)}_{JT,33}(-\tau,\cos\theta)
&=
-
H^{(Q)}_{JT,33}(\tau,\cos\theta),
\\
H^{(-Q)}_{JT,0}(-\tau,\cos\theta)
&=
-
H^{(Q)}_{JT,0}(\tau,\cos\theta).
\end{align}

\paragraph{Tensor-tensor}

\begin{align}
H^{(-Q)}_{TT,2233}(-\tau,\cos\theta)
&=
H^{(Q)}_{TT,2233}(\tau,\cos\theta),
\\
H^{(-Q)}_{TT,2\delta3}(-\tau,\cos\theta)
&=
H^{(Q)}_{TT,2\delta3}(\tau,\cos\theta),
\\
H^{(-Q)}_{TT,\delta\delta}(-\tau,\cos\theta)
&=
H^{(Q)}_{TT,\delta\delta}(\tau,\cos\theta),
\\
H^{(-Q)}_{TT,33}(-\tau,\cos\theta)
&=
H^{(Q)}_{TT,33}(\tau,\cos\theta),
\\
H^{(-Q)}_{TT,22}(-\tau,\cos\theta)
&=
H^{(Q)}_{TT,22}(\tau,\cos\theta),
\\
H^{(-Q)}_{TT,23}(-\tau,\cos\theta)
&=
H^{(Q)}_{TT,23}(\tau,\cos\theta),
\\
H^{(-Q)}_{TT,\delta}(-\tau,\cos\theta)
&=
H^{(Q)}_{TT,\delta}(\tau,\cos\theta),
\\
H^{(-Q)}_{TT,0}(-\tau,\cos\theta)
&=
H^{(Q)}_{TT,0}(\tau,\cos\theta),
\end{align}
and
\begin{align}
H^{(-Q)}_{TT,223}(-\tau,\cos\theta)
&=
-
H^{(Q)}_{TT,223}(\tau,\cos\theta),
\\
H^{(-Q)}_{TT,2\delta}(-\tau,\cos\theta)
&=
-
H^{(Q)}_{TT,2\delta}(\tau,\cos\theta),
\\
H^{(-Q)}_{TT,233}(-\tau,\cos\theta)
&=
-
H^{(Q)}_{TT,233}(\tau,\cos\theta),
\\
H^{(-Q)}_{TT,\delta3}(-\tau,\cos\theta)
&=
-
H^{(Q)}_{TT,\delta3}(\tau,\cos\theta),
\\
H^{(-Q)}_{TT,3}(-\tau,\cos\theta)
&=
-
H^{(Q)}_{TT,3}(\tau,\cos\theta),
\\
H^{(-Q)}_{TT,2}(-\tau,\cos\theta)
&=
-
H^{(Q)}_{TT,2}(\tau,\cos\theta).
\end{align}

\subsection{Crossing equations for the \(h\)-functions \label{app:Crossingh-functions}}

The crossing equations for the \(h\)-functions are a slight modification of those for the \(H\)-basis functions. In defining the normalized NLO functions \(h\), we factored out the leading Ward-identity factors in all correlators involving the current \(J\). Since the charge changes sign under \(Q\to -Q\), each insertion of \(J\) produces an additional minus sign. In the mixed charged-scalar channels \(SJ\) and \(ST\), there is one more sign: the normalization factor
\begin{equation}
\kappa_{SJ}=\frac{q}{2Q},  \hfill  \kappa_{ST} = \frac{3 q}{4Q},
\end{equation}
is odd under \(Q\to -Q\). As a result, the crossing parities of the charged \(SJ\) and \(ST\) \(h\)-functions are flipped relative to the corresponding neutral-scalar case.

The \(h\)-functions do not carry an explicit superscript \(Q\). They are defined after factoring out the leading charge-dependent factors, so the remaining crossing equations are written directly as parity conditions under \(\tau\to -\tau\). In the scalar-scalar sector, we also use the redefined function introduced in \eqref{eq:hSSRedefinition}, which absorbs the inhomogeneous term coming from the mismatch of the external large-charge dimensions.

\paragraph{Scalar-scalar}

\begin{equation}
h_{SS}(-\tau,\cos\theta)
=
h_{SS}(\tau,\cos\theta).
\end{equation}

\paragraph{Scalar-current}

\begin{equation}
h_{SJ,0}(-\tau,\cos\theta)
=
-
h_{SJ,0}(\tau,\cos\theta),
\end{equation}
and
\begin{equation}
h_{SJ,3}(-\tau,\cos\theta)
=
h_{SJ,3}(\tau,\cos\theta).
\end{equation}

\paragraph{Current-current}

\begin{align}
h_{JJ,23}(-\tau,\cos\theta)
&=
h_{JJ,23}(\tau,\cos\theta),
\\
h_{JJ,\delta}(-\tau,\cos\theta)
&=
h_{JJ,\delta}(\tau,\cos\theta),
\\
h_{JJ,0}(-\tau,\cos\theta)
&=
h_{JJ,0}(\tau,\cos\theta),
\end{align}
and
\begin{align}
h_{JJ,3}(-\tau,\cos\theta)
&=
-
h_{JJ,3}(\tau,\cos\theta),
\\
h_{JJ,2}(-\tau,\cos\theta)
&=
-
h_{JJ,2}(\tau,\cos\theta).
\end{align}

\paragraph{Scalar-tensor}

\begin{align}
h_{ST,33}(-\tau,\cos\theta)
&=
-
h_{ST,33}(\tau,\cos\theta),
\\
h_{ST,0}(-\tau,\cos\theta)
&=
-
h_{ST,0}(\tau,\cos\theta),
\end{align}
and
\begin{equation}
h_{ST,3}(-\tau,\cos\theta)
=
h_{ST,3}(\tau,\cos\theta).
\end{equation}

\paragraph{Current-tensor}

\begin{align}
h_{JT,23}(-\tau,\cos\theta)
&=
h_{JT,23}(\tau,\cos\theta),
\\
h_{JT,\delta}(-\tau,\cos\theta)
&=
h_{JT,\delta}(\tau,\cos\theta),
\\
h_{JT,33}(-\tau,\cos\theta)
&=
h_{JT,33}(\tau,\cos\theta),
\\
h_{JT,0}(-\tau,\cos\theta)
&=
h_{JT,0}(\tau,\cos\theta),
\end{align}
and
\begin{align}
h_{JT,233}(-\tau,\cos\theta)
&=
-
h_{JT,233}(\tau,\cos\theta),
\\
h_{JT,\delta3}(-\tau,\cos\theta)
&=
-
h_{JT,\delta3}(\tau,\cos\theta),
\\
h_{JT,3}(-\tau,\cos\theta)
&=
-
h_{JT,3}(\tau,\cos\theta),
\\
h_{JT,2}(-\tau,\cos\theta)
&=
-
h_{JT,2}(\tau,\cos\theta).
\end{align}

\paragraph{Tensor-tensor}

\begin{align}
h_{TT,2233}(-\tau,\cos\theta)
&=
h_{TT,2233}(\tau,\cos\theta),
\\
h_{TT,2\delta3}(-\tau,\cos\theta)
&=
h_{TT,2\delta3}(\tau,\cos\theta),
\\
h_{TT,\delta\delta}(-\tau,\cos\theta)
&=
h_{TT,\delta\delta}(\tau,\cos\theta),
\\
h_{TT,33}(-\tau,\cos\theta)
&=
h_{TT,33}(\tau,\cos\theta),
\\
h_{TT,22}(-\tau,\cos\theta)
&=
h_{TT,22}(\tau,\cos\theta),
\\
h_{TT,23}(-\tau,\cos\theta)
&=
h_{TT,23}(\tau,\cos\theta),
\\
h_{TT,\delta}(-\tau,\cos\theta)
&=
h_{TT,\delta}(\tau,\cos\theta),
\\
h_{TT,0}(-\tau,\cos\theta)
&=
h_{TT,0}(\tau,\cos\theta),
\end{align}
and
\begin{align}
h_{TT,223}(-\tau,\cos\theta)
&=
-
h_{TT,223}(\tau,\cos\theta),
\\
h_{TT,2\delta}(-\tau,\cos\theta)
&=
-
h_{TT,2\delta}(\tau,\cos\theta),
\\
h_{TT,233}(-\tau,\cos\theta)
&=
-
h_{TT,233}(\tau,\cos\theta),
\\
h_{TT,\delta3}(-\tau,\cos\theta)
&=
-
h_{TT,\delta3}(\tau,\cos\theta),
\\
h_{TT,3}(-\tau,\cos\theta)
&=
-
h_{TT,3}(\tau,\cos\theta),
\\
h_{TT,2}(-\tau,\cos\theta)
&=
-
h_{TT,2}(\tau,\cos\theta).
\end{align}

\section{Contact singularities from the large-charge EFT}
\label{app:EFTContactSingularity}

In this appendix we illustrate explicitly how contact singularities arise in the large-charge EFT. We consider the correlator of two charged scalar probes in \(d=3\). At the order relevant for the discussion in the main text, its singular part is controlled by the propagator of the Goldstone mode. Near coincident points on the cylinder,
\begin{equation}
\tau=\t=0,
\end{equation}
the cylinder is locally flat and the propagator behaves as
\begin{equation}
\label{eq:GoldstonePropagatorLocal}
G(\tau,\t)
\sim
\f{1}{\sqrt{\tau^2+\t^2}},
\end{equation}
up to an overall numerical normalization. Differentiating \eqref{eq:GoldstonePropagatorLocal} with respect to \(\tau\), we find
\begin{equation}
\label{eq:GoldstonePropagatorDerivative}
\partial_\tau G(\tau,\t)
\sim
-\f{\tau}{(\tau^2+\t^2)^{3/2}}.
\end{equation}
Consequently, at \(\tau=\eps>0\),
\begin{equation}
\label{eq:GoldstoneDeltaSequence}
-\partial_\tau G(\tau,\t)\Big|_{\tau=\eps}
\sim
\f{\eps}{(\eps^2+\t^2)^{3/2}} \underset{\eps\to 0}{\sim} \delta_{S^2}(\vec n),
\end{equation}
which can be shown by integrating it with a smooth test function \(f(\t)\).

Alternatively, the same conclusion follows directly from the spectral representation of the propagator on the cylinder.
\begin{equation}
\label{eq:GoldstoneSpectralPropagator}
G(\tau,\vec n,\vec n_0)
=
-\f{|\tau|}{8\pi}
+
\sum_{\ell=1}^{\infty}
\sum_{m=-\ell}^{\ell}
\f{1}{2\omega_\ell}
e^{-\omega_\ell|\tau|}
Y_{\ell m}(\vec n)
Y_{\ell m}^*(\vec n_0).
\end{equation}
For the derivative we get
\begin{equation}
\label{eq:GoldstoneDerivativeDiscontinuityDefinition}
\lim_{\eps\to0^+}
\partial_\tau G(\mp\eps,\vec n,\vec n_0)
=
\pm \f{1}{2}
\sum_{\ell=0}^{\infty}
\sum_{m=-\ell}^{\ell}
Y_{\ell m}(\vec n)
Y_{\ell m}^*(\vec n_0)
=
\pm \f{1}{2}
\delta_{S^2}(\vec n,\vec n_0),
\end{equation}
which reproduces precisely the contact distribution obtained in the flat-space limit.

\section{Full set of crossing equations}
\label{app:full-crossing-equations}

In this appendix we list the crossing equations obtained directly from the individual \(h\)-functions before eliminating dependent equations. Overall numerical constants have been absorbed into the definition of the polynomials on the right-hand side. The main text uses the reduced independent subset of these equations. Throughout this appendix,
\begin{equation}
n=1,2,\ldots .
\end{equation}
For \(\ell\geq2\), the primary states are organized into a finite number of Regge trajectories labelled by \(i\). At \(\ell=0\) and \(\ell=1\), the sums run over the physical primary states contributing at the order considered. If no equation is displayed for a given structure at a particular spin, this means that the corresponding structure does not exist at that spin.

\paragraph{Scalar-scalar}

For \(\ell\geq2\),
\begin{equation}
\sum_i
|\lambda_{S,\ell,i}|^2
\omega_{\ell,i}^{2n-1}
=
P^{(SS)}_{n-1}(J_\ell^2).
\end{equation}

For \(\ell=1\),
\begin{equation}
1
+
\sum_i
|\lambda_{S,1,i}|^2
\omega_{1,i}^{2n-1}
=
P^{(SS)}_{n-1}(1).
\end{equation}

For \(\ell=0\),
\begin{equation}
\delta_{n,1}
+
\sum_i
|\lambda_{S,0,i}|^2
\omega_{0,i}^{2n-1}
=
P^{(SS)}_{n-1}(0).
\end{equation}

\paragraph{Scalar-current}

For \(\ell\geq2\),
\begin{align}
\sum_i
\lambda_{S,\ell,i}^*\lambda_{J,\ell,i}
\frac{\omega_{\ell,i}^{2n}}{J_\ell^2}
&=
P^{(SJ,3)}_{n-1}(J_\ell^2),
\\
\sum_i
\lambda_{S,\ell,i}^*\lambda_{J,\ell,i}
\omega_{\ell,i}^{2n-2}
&=
P^{(SJ,0)}_{n-1}(J_\ell^2).
\end{align}

For \(\ell=1\),
\begin{align}
1
+
\sum_i
\lambda_{S,1,i}^*\lambda_{J,1,i}
\omega_{1,i}^{2n}
&=
P^{(SJ,3)}_{n-1}(1),
\\
1
+
\sum_i
\lambda_{S,1,i}^*\lambda_{J,1,i}
\omega_{1,i}^{2n-2}
&=
P^{(SJ,0)}_{n-1}(1).
\end{align}

\paragraph{Current-current}

For \(\ell\geq2\),
\begin{align}
\sum_i
|\lambda_{J,\ell,i}|^2
\frac{\omega_{\ell,i}^{2n+1}}{J_\ell^4}
&=
P^{(JJ,23)}_{n-1}(J_\ell^2),
\\
\sum_i
|\lambda_{J,\ell,i}|^2
\frac{\omega_{\ell,i}^{2n+1}}{J_\ell^4}
&=
P^{(JJ,\delta)}_{n-1}(J_\ell^2),
\\
\sum_i
|\lambda_{J,\ell,i}|^2
\omega_{\ell,i}^{2n-1}
&=
P^{(JJ,0)}_{n}(J_\ell^2),
\\
\sum_i
|\lambda_{J,\ell,i}|^2
\frac{\omega_{\ell,i}^{2n-1}}{J_\ell^2}
&=
P^{(JJ,3)}_{n-1}(J_\ell^2),
\\
\sum_i
|\lambda_{J,\ell,i}|^2
\frac{\omega_{\ell,i}^{2n-1}}{J_\ell^2}
&=
P^{(JJ,2)}_{n-1}(J_\ell^2).
\end{align}

For \(\ell=1\),
\begin{align}
1
+
\sum_i
|\lambda_{J,1,i}|^2
\omega_{1,i}^{2n+1}
&=
P^{(JJ,\delta)}_{n-1}(1),
\\
1
+
\sum_i
|\lambda_{J,1,i}|^2
\omega_{1,i}^{2n-1}
&=
P^{(JJ,0)}_{n}(1),
\\
1
+
\sum_i
|\lambda_{J,1,i}|^2
\omega_{1,i}^{2n-1}
&=
P^{(JJ,3)}_{n-1}(1),
\\
1
+
\sum_i
|\lambda_{J,1,i}|^2
\omega_{1,i}^{2n-1}
&=
P^{(JJ,2)}_{n-1}(1).
\end{align}

\paragraph{Scalar-tensor}

For \(\ell\geq2\),
\begin{align}
\sum_i
\lambda_{S,\ell,i}^*\lambda_{T,\ell,i}
\frac{J_\ell^2-\omega_{\ell,i}^2}
{J_\ell^2\left(J_\ell^2-1\right)}
\omega_{\ell,i}^{2n-2}
&=
P^{(ST,33)}_{n-2}(J_\ell^2),
\\
\sum_i
\lambda_{S,\ell,i}^*\lambda_{T,\ell,i}
\omega_{\ell,i}^{2n-2}
&=
P^{(ST,0)}_{n-1}(J_\ell^2),
\\
\sum_i
\lambda_{S,\ell,i}^*\lambda_{T,\ell,i}
\frac{\omega_{\ell,i}^{2n}}{J_\ell^2}
&=
P^{(ST,3)}_{n-1}(J_\ell^2).
\end{align}

For \(\ell=1\),
\begin{equation}
P^{(ST,0)}_{n-1}(1)
=
P^{(ST,3)}_{n-1}(1)
=
1+\sum_i\lambda_{S,1,i}^*\lambda_{T,1,i},
\qquad
\omega_{1,i}=1.
\end{equation}

\paragraph{Current-tensor}

For \(\ell\geq2\),
\begin{align}
\sum_i
\lambda_{J,\ell,i}^*\lambda_{T,\ell,i}
\frac{\omega_{\ell,i}^{2n+1}}{J_\ell^4}
&=
P^{(JT,23)}_{n-1}(J_\ell^2),
\\
\sum_i
\lambda_{J,\ell,i}^*\lambda_{T,\ell,i}
\frac{\omega_{\ell,i}^{2n+1}}{J_\ell^4}
&=
P^{(JT,\delta)}_{n-1}(J_\ell^2),
\\
\sum_i
\lambda_{J,\ell,i}^*\lambda_{T,\ell,i}
\frac{J_\ell^2-\omega_{\ell,i}^2}
{J_\ell^2\left(J_\ell^2-1\right)}
\omega_{\ell,i}^{2n-1}
&=
P^{(JT,33)}_{n-1}(J_\ell^2),
\\
\sum_i
\lambda_{J,\ell,i}^*\lambda_{T,\ell,i}
\omega_{\ell,i}^{2n-1}
&=
P^{(JT,0)}_{n}(J_\ell^2),
\\
\sum_i
\lambda_{J,\ell,i}^*\lambda_{T,\ell,i}
\frac{\left(J_\ell^2-\omega_{\ell,i}^2\right)\omega_{\ell,i}^{2n-1}}
{J_\ell^4\left(J_\ell^2-1\right)}
&=
P^{(JT,233)}_{n-2}(J_\ell^2),
\qquad
\ell\geq3,
\\
\sum_i
\lambda_{J,\ell,i}^*\lambda_{T,\ell,i}
\frac{\left(J_\ell^2-\omega_{\ell,i}^2\right)\omega_{\ell,i}^{2n-1}}
{J_\ell^4\left(J_\ell^2-1\right)}
&=
P^{(JT,\delta3)}_{n-2}(J_\ell^2),
\\
\sum_i
\lambda_{J,\ell,i}^*\lambda_{T,\ell,i}
\frac{\omega_{\ell,i}^{2n-1}}{J_\ell^2}
&=
P^{(JT,3)}_{n-1}(J_\ell^2),
\\
\sum_i
\lambda_{J,\ell,i}^*\lambda_{T,\ell,i}
\frac{\omega_{\ell,i}^{2n-1}}{J_\ell^2}
&=
P^{(JT,2)}_{n-1}(J_\ell^2).
\end{align}

For \(\ell=1\),
\begin{equation}
P^{(JT,\delta)}_{n-1}(1)
=
P^{(JT,0)}_{n}(1)
=
P^{(JT,3)}_{n-1}(1)
=
P^{(JT,2)}_{n-1}(1)
=
1+\sum_i\lambda_{J,1,i}^*\lambda_{T,1,i},
\qquad
\omega_{1,i}=1.
\end{equation}

\paragraph{Tensor-tensor}

For \(\ell\geq2\),
\begin{align}
\sum_i
|\lambda_{T,\ell,i}|^2
\frac{\left(J_\ell^2-\omega_{\ell,i}^2\right)^2}
{J_\ell^4\left(J_\ell^2-1\right)^2}
\omega_{\ell,i}^{2n-1}
&=
P^{(TT,2233)}_{n-2}(J_\ell^2),
\qquad
\ell\geq4,
\\
\sum_i
|\lambda_{T,\ell,i}|^2
\frac{\left(J_\ell^2-\omega_{\ell,i}^2\right)^2}
{J_\ell^4\left(J_\ell^2-1\right)^2}
\omega_{\ell,i}^{2n-1}
&=
P^{(TT,2\delta3)}_{n-2}(J_\ell^2),
\qquad
\ell\geq3,
\\
\sum_i
|\lambda_{T,\ell,i}|^2
\frac{\left(J_\ell^2-\omega_{\ell,i}^2\right)^2}
{J_\ell^4\left(J_\ell^2-1\right)^2}
\omega_{\ell,i}^{2n-1}
&=
P^{(TT,\delta\delta)}_{n-2}(J_\ell^2),
\\
\sum_i
|\lambda_{T,\ell,i}|^2
\frac{J_\ell^2-\omega_{\ell,i}^2}
{J_\ell^2\left(J_\ell^2-1\right)}
\omega_{\ell,i}^{2n-1}
&=
P^{(TT,33)}_{n-1}(J_\ell^2),
\\
\sum_i
|\lambda_{T,\ell,i}|^2
\frac{J_\ell^2-\omega_{\ell,i}^2}
{J_\ell^2\left(J_\ell^2-1\right)}
\omega_{\ell,i}^{2n-1}
&=
P^{(TT,22)}_{n-1}(J_\ell^2),
\\
\sum_i
|\lambda_{T,\ell,i}|^2
\frac{J_\ell^2-\omega_{\ell,i}^2}
{J_\ell^2\left(J_\ell^2-1\right)}
\omega_{\ell,i}^{2n-1}
&=
P^{(TT,23)}_{n-1}(J_\ell^2),
\\
\sum_i
|\lambda_{T,\ell,i}|^2
\frac{\omega_{\ell,i}^{2n+1}}{J_\ell^4}
&=
P^{(TT,\delta)}_{n-1}(J_\ell^2),
\\
\sum_i
|\lambda_{T,\ell,i}|^2
\omega_{\ell,i}^{2n-1}
&=
P^{(TT,0)}_{n}(J_\ell^2),
\\
\sum_i
|\lambda_{T,\ell,i}|^2
\frac{\left(J_\ell^2-\omega_{\ell,i}^2\right)\omega_{\ell,i}^{2n-1}}
{J_\ell^4\left(J_\ell^2-1\right)}
&=
P^{(TT,223)}_{n-2}(J_\ell^2),
\qquad
\ell\geq3,
\\
\sum_i
|\lambda_{T,\ell,i}|^2
\frac{\left(J_\ell^2-\omega_{\ell,i}^2\right)\omega_{\ell,i}^{2n-1}}
{J_\ell^4\left(J_\ell^2-1\right)}
&=
P^{(TT,2\delta)}_{n-2}(J_\ell^2),
\\
\sum_i
|\lambda_{T,\ell,i}|^2
\frac{\left(J_\ell^2-\omega_{\ell,i}^2\right)\omega_{\ell,i}^{2n-1}}
{J_\ell^4\left(J_\ell^2-1\right)}
&=
P^{(TT,233)}_{n-2}(J_\ell^2),
\qquad
\ell\geq3,
\\
\sum_i
|\lambda_{T,\ell,i}|^2
\frac{\left(J_\ell^2-\omega_{\ell,i}^2\right)\omega_{\ell,i}^{2n-1}}
{J_\ell^4\left(J_\ell^2-1\right)}
&=
P^{(TT,\delta3)}_{n-2}(J_\ell^2),
\\
\sum_i
|\lambda_{T,\ell,i}|^2
\frac{\omega_{\ell,i}^{2n-1}}{J_\ell^2}
&=
P^{(TT,3)}_{n-1}(J_\ell^2),
\\
\sum_i
|\lambda_{T,\ell,i}|^2
\frac{\omega_{\ell,i}^{2n-1}}{J_\ell^2}
&=
P^{(TT,2)}_{n-1}(J_\ell^2).
\end{align}

For \(\ell=1\),
\begin{equation}
P^{(TT,\delta)}_{n-1}(1)
=
P^{(TT,0)}_{n}(1)
=
P^{(TT,3)}_{n-1}(1)
=
P^{(TT,2)}_{n-1}(1)
=
1+\sum_i|\lambda_{T,1,i}|^2,
\qquad
\omega_{1,i}=1.
\end{equation}
In the isolated spin-one equations above, the sums run over all spin-one primaries with excitation energy \(\omega_{1,i}=1\) that can couple to the corresponding pair of probes. Their contributions have the same dependence on \(\tau\) and \(\theta\) as the first descendant of the ground-state primary. If the stronger non-degeneracy assumption, which excludes degeneracy between the first descendant of the ground-state primary and the new spin-one primaries entering at this order, is imposed, these sums vanish.

Whenever the index of a polynomial is negative, the corresponding polynomial is identically zero. In particular, all equations involving \(P_{n-2}\) have a vanishing right-hand side at \(n=1\).

\section{Decoupling of non-Goldstone trajectories from the current}
\label{app:SV-current-decoupling}

In this appendix we prove the statement used in Section~\ref{subsec:SV-sector} that under the assumption
\begin{equation}
x_i(0)\neq 0,
\qquad
i=2,\ldots,N,
\end{equation}
the non-Goldstone trajectories do not contribute to the current-current channel. We present two equivalent arguments. The first uses the recurrence relation and divisibility by powers of \(w\). The second uses the generating function.

\subsection{Proof from divisibility and the recurrence relation}

Subtract the Goldstone contribution from the \(VV\) equation and define
\begin{equation}
R_n^{(VV)}(w)
\equiv
P_n^{(VV)}(w)
-
\frac{p_{VT}^2}{p_{TT}}\,w^n .
\end{equation}
Using the spectral representation, this is
\begin{equation}
R_n^{(VV)}(w)
=
\sum_{i=2}^{N}
V_i(w)\,x_i(w)^n .
\end{equation}
The polynomial consistency conditions imply
\begin{equation}
R_1^{(VV)}(w)
=
w
\left(
p_{VV}
-
\frac{p_{VT}^2}{p_{TT}}
\right),
\end{equation}
and, for \(n\geq 2\),
\begin{equation}
R_n^{(VV)}(w)
=
w^2\,\widetilde R_{n-2}^{(VV)}(w),
\end{equation}
with \(\widetilde R_{n-2}^{(VV)}(w)\) polynomial in \(w\).

The sequence \(R_n^{(VV)}\) is a finite sum over the non-Goldstone roots \(x_2,\ldots,x_N\). Therefore it satisfies the recurrence relation with characteristic roots \(x_2,\ldots,x_N\). Let
\begin{equation}
q_k(w),
\qquad
k=1,\ldots,N-1,
\end{equation}
be the elementary symmetric polynomials in \(x_2,\ldots,x_N\). Then
\begin{align}
R_{n+N-1}^{(VV)}
&=
q_1 R_{n+N-2}^{(VV)}
-
q_2 R_{n+N-3}^{(VV)}
+\cdots
+
(-1)^N q_{N-1} R_n^{(VV)} .
\end{align}
Set \(n=1\). All terms \(R_m^{(VV)}\) with \(m\geq 2\) are divisible by \(w^2\). Hence the recurrence implies
\begin{equation}
q_{N-1}(w)\,R_1^{(VV)}(w)
=
O(w^2).
\end{equation}
But
\begin{equation}
q_{N-1}(w)
=
\prod_{i=2}^{N}x_i(w).
\end{equation}
By assumption,
\begin{equation}
q_{N-1}(0)\neq 0.
\end{equation}
Therefore \(R_1^{(VV)}(w)\) itself must be divisible by \(w^2\). Since
\begin{equation}
R_1^{(VV)}(w)
=
w
\left(
p_{VV}
-
\frac{p_{VT}^2}{p_{TT}}
\right),
\end{equation}
we conclude
\begin{equation}
p_{VV}
=
\frac{p_{VT}^2}{p_{TT}}.
\end{equation}
Thus
\begin{equation}
R_1^{(VV)}(w)=0.
\end{equation}
For \(w\geq 3\),
\begin{equation}
R_1^{(VV)}(w)
=
\sum_{i=2}^{N}
V_i(w)\,x_i(w).
\end{equation}
Using positivity in the diagonal \(VV\) channel and positivity of the excitation energies,
\begin{equation}
V_i(w)\geq 0,
\qquad
x_i(w)>0,
\end{equation}
we obtain
\begin{equation}
V_i(w)=0,
\qquad
i=2,\ldots,N .
\end{equation}
This proves the claim.

\subsection{Generating-function proof}

The same conclusion can be reached using the generating function. Define
\begin{equation}
R^{(VV)}(t,w)
=
\sum_{n=1}^{\infty}
R_n^{(VV)}(w)t^n .
\end{equation}
From the divisibility conditions above,
\begin{equation}
R^{(VV)}(t,w)
=
w
\left(
p_{VV}
-
\frac{p_{VT}^2}{p_{TT}}
\right)t
+
O(w^2).
\end{equation}
In particular, the coefficient of order \(w\) is a polynomial in \(t\), proportional only to \(t\). On the other hand, the spectral representation gives
\begin{equation}
R^{(VV)}(t,w)
=
\sum_{i=2}^{N}
\frac{V_i(w)\,x_i(w)t}{1-tx_i(w)} .
\end{equation}
If
\begin{equation}
x_i(0)\neq 0,
\qquad
i=2,\ldots,N,
\end{equation}
then near \(w=0\) the possible poles in \(t\) are located at
\begin{equation}
t=\frac{1}{x_i(0)} .
\end{equation}
But the order-\(w\) term obtained from the consistency conditions has no such poles: it is simply proportional to \(t\). Therefore the residues of the spectral representation at these poles must vanish. This forces the order-\(w\) non-Goldstone current contribution to vanish, and gives again
\begin{equation}
p_{VV}
=
\frac{p_{VT}^2}{p_{TT}} .
\end{equation}
Then
\begin{equation}
R_1^{(VV)}(w)=0,
\end{equation}
which reproduces the result of the divisibility proof.

\bibliographystyle{JHEP}
\bibliography{BootstrapSpinning}{}

@article{Costa:2011mg,
    author = "Costa, Miguel S. and Penedones, Joao and Poland, David and Rychkov, Slava",
    title = "{Spinning Conformal Correlators}",
    eprint = "1107.3554",
    archivePrefix = "arXiv",
    primaryClass = "hep-th",
    reportNumber = "LPTENS-11-22, NSF-KITP-11-128",
    doi = "10.1007/JHEP11(2011)071",
    journal = "JHEP",
    volume = "11",
    pages = "071",
    year = "2011"
}

@article{Costa:2011dw,
    author = "Costa, Miguel S. and Penedones, Joao and Poland, David and Rychkov, Slava",
    title = "{Spinning Conformal Blocks}",
    eprint = "1109.6321",
    archivePrefix = "arXiv",
    primaryClass = "hep-th",
    reportNumber = "LPTENS-11-37",
    doi = "10.1007/JHEP11(2011)154",
    journal = "JHEP",
    volume = "11",
    pages = "154",
    year = "2011"
}

@article{Simmons-Duffin:2016wlq,
    author = "Simmons-Duffin, David",
    title = "{The Lightcone Bootstrap and the Spectrum of the 3d Ising CFT}",
    eprint = "1612.08471",
    archivePrefix = "arXiv",
    primaryClass = "hep-th",
    doi = "10.1007/JHEP03(2017)086",
    journal = "JHEP",
    volume = "03",
    pages = "086",
    year = "2017"
}

@article{Fitzpatrick:2012yx,
    author = "Fitzpatrick, A. Liam and Kaplan, Jared and Poland, David and Simmons-Duffin, David",
    title = "{The Analytic Bootstrap and AdS Superhorizon Locality}",
    eprint = "1212.3616",
    archivePrefix = "arXiv",
    primaryClass = "hep-th",
    doi = "10.1007/JHEP12(2013)004",
    journal = "JHEP",
    volume = "12",
    pages = "004",
    year = "2013"
}

@article{Komargodski:2012ek,
    author = "Komargodski, Zohar and Zhiboedov, Alexander",
    title = "{Convexity and Liberation at Large Spin}",
    eprint = "1212.4103",
    archivePrefix = "arXiv",
    primaryClass = "hep-th",
    doi = "10.1007/JHEP11(2013)140",
    journal = "JHEP",
    volume = "11",
    pages = "140",
    year = "2013"
}

@article{Hellerman:2015nra,
    author = "Hellerman, Simeon and Orlando, Domenico and Reffert, Susanne and Watanabe, Masataka",
    title = "{On the CFT Operator Spectrum at Large Global Charge}",
    eprint = "1505.01537",
    archivePrefix = "arXiv",
    primaryClass = "hep-th",
    doi = "10.1007/JHEP12(2015)071",
    journal = "JHEP",
    volume = "12",
    pages = "071",
    year = "2015"
}

@article{Monin:2016jmo,
    author = "Monin, Alexander and Pirtskhalava, David and Rattazzi, Riccardo and Seibold, Fiona K.",
    title = "{Semiclassics, Goldstone Bosons and CFT data}",
    eprint = "1611.02912",
    archivePrefix = "arXiv",
    primaryClass = "hep-th",
    doi = "10.1007/JHEP06(2017)011",
    journal = "JHEP",
    volume = "06",
    pages = "011",
    year = "2017"
}

@article{Dondi:2022zna,
    author = "Dondi, Nicola and Hellerman, Simeon and Kalogerakis, Ioannis and Moser, Rafael and Orlando, Domenico and Reffert, Susanne",
    title = "{Fermionic CFTs at large charge and large N}",
    eprint = "2211.15318",
    archivePrefix = "arXiv",
    primaryClass = "hep-th",
    doi = "10.1007/JHEP08(2023)180",
    journal = "JHEP",
    volume = "08",
    pages = "180",
    year = "2023"
}

@article{Jafferis:2017zna,
    author = "Jafferis, Daniel and Mukhametzhanov, Baur and Zhiboedov, Alexander",
    title = "{Conformal Bootstrap At Large Charge}",
    eprint = "1710.11161",
    archivePrefix = "arXiv",
    primaryClass = "hep-th",
    doi = "10.1007/JHEP05(2018)043",
    journal = "JHEP",
    volume = "05",
    pages = "043",
    year = "2018"
}

@article{Kiaee:2025jly,
    author = "Kiaee, Kasra and Monin, Alexander",
    title = "{Large charge bootstrap with U(1) current probes}",
    eprint = "2512.20803",
    archivePrefix = "arXiv",
    primaryClass = "hep-th",
    doi = "10.1007/JHEP05(2026)121",
    journal = "JHEP",
    volume = "05",
    pages = "121",
    year = "2026"
}

@article{Caron-Huot:2017vep,
    author = "Caron-Huot, Simon",
    title = "{Analyticity in Spin in Conformal Theories}",
    eprint = "1703.00278",
    archivePrefix = "arXiv",
    primaryClass = "hep-th",
    doi = "10.1007/JHEP09(2017)078",
    journal = "JHEP",
    volume = "09",
    pages = "078",
    year = "2017"
}

@article{Esposito:2017qpj,
    author = "Esposito, A. and Garcia-Saenz, S. and Nicolis, A. and Penco, R.",
    title = "{Conformal solids and holography}",
    eprint = "1708.09391",
    archivePrefix = "arXiv",
    primaryClass = "hep-th",
    doi = "10.1007/JHEP12(2017)113",
    journal = "JHEP",
    volume = "12",
    pages = "113",
    year = "2017"
}

@article{Cuomo:2021qws,
    author = "Cuomo, Gabriel and Delacretaz, Luca V. and Mehta, Umang",
    title = "{Large Charge Sector of 3d Parity-Violating CFTs}",
    eprint = "2102.05046",
    archivePrefix = "arXiv",
    primaryClass = "hep-th",
    reportNumber = "EFI-21-1",
    doi = "10.1007/JHEP05(2021)115",
    journal = "JHEP",
    volume = "05",
    pages = "115",
    year = "2021"
}

\end{document}